\documentclass[10pt,journal,compsoc]{IEEEtran}

\ifCLASSOPTIONcompsoc
\usepackage[nocompress]{cite}
\else
\usepackage{cite}
\fi
\usepackage{booktabs} 
\usepackage{xcolor}
\usepackage{soul}
\usepackage{epsfig}
\usepackage{graphicx}
\usepackage{amsmath}
\usepackage{amssymb}
\usepackage{courier}
\usepackage{helvet}
\usepackage{courier}
\usepackage{algorithm}
\usepackage{algorithmic}
\usepackage{multirow}
\usepackage{url}
\usepackage{booktabs}
\usepackage{array}
\usepackage{paralist,algorithmic,algorithm}
\usepackage{balance}
\usepackage[numbers]{natbib}
\usepackage{makecell}

\newcommand{\jj}{{\bf j}}
\newcommand{\e}{{\bf e}}

\newcommand{\x}{{\bf x}}
\newcommand{\y}{{\bf y}}

\newcommand{\uu}{{\bf u}}
\newcommand{\s}{{\bf s}}

\newcommand{\N}{\mathcal{N}}

\newcommand{\R}{\mathbb{R}}

\newcommand{\D}{\mathcal{D}}
\newtheorem{defn}{Definition}

\ifCLASSINFOpdf
\else
\fi

\begin{document}

\title{JobFormer: Skill-Aware Job Recommendation with Semantic-Enhanced Transformer}


\author{ZhiHao Guan,~\IEEEmembership{}
	    Jia-Qi Yang,~\IEEEmembership{}
	    Yang Yang,~\IEEEmembership{Member, IEEE}
	    Hengshu Zhu,~\IEEEmembership{Senior Member, IEEE}
	    Wenjie Li,~\IEEEmembership{}
        and Hui Xiong,~\IEEEmembership{Fellow, IEEE}
\IEEEcompsocitemizethanks{\IEEEcompsocthanksitem Zhihao Guan and Yang Yang are with the Nanjing University of Science and Technology, Nanjing 210094, China. E-mail: {zhguan,yyang}@njust.edu.cn}
\IEEEcompsocitemizethanks{\IEEEcompsocthanksitem Jia-Qi Yang is with State Key Laboratory for Novel Software Technology, Nanjing University, Nanjing 210023, China. E-mail:yangjq@lamda.nju.edu.com}
\IEEEcompsocitemizethanks{\IEEEcompsocthanksitem Hengshu Zhu is with Career Science Lab, BOSS Zhipin, Beijing 100028, China. E-mail:zhuhengshu@gmail.com}
\IEEEcompsocitemizethanks{\IEEEcompsocthanksitem Wenjie Li is with the Hong Kong Polytechnic University, Hong Kong, China. E-mail: wenjie.li@polyu.edu.hk.}
\IEEEcompsocitemizethanks{\IEEEcompsocthanksitem  Hui Xiong is with the Artificial Intelligence Thrust, The Hong Kong University of Science and Technology, Guangzhou, China. E-mail: xionghui@ust.hk}
\thanks{Yang Yang (Corresponding Author) is with PCA Lab, Key Lab of Intelligent Perception and Systems for High-Dimensional Information of Ministry of Education, and Jiangsu Key Lab of Image and Video Understanding for Social Security, School of Computer Science and Engineering, Nanjing University of Science and Technology.}}


\IEEEtitleabstractindextext{%

\begin{abstract}
Job recommendation aims to provide potential talents with suitable job descriptions (JDs) consistent with their career trajectory, which plays an essential role in proactive talent recruitment. In real-world management scenarios, the available JD-user records always consist of JDs, user profiles, and click data, in which the user profiles are typically summarized as the user's skill distribution for privacy reasons. Although existing sophisticated recommendation methods can be directly employed, effective recommendation still has challenges considering the information deficit of JD itself and the natural heterogeneous gap between JD and user profile. To address these challenges, we proposed a novel skill-aware recommendation model based on the designed semantic-enhanced transformer to parse JDs and complete personalized job recommendation. Specifically, we first model the relative items of each JD and then adopt an encoder with the local-global attention mechanism to better mine the intra-job and inter-job dependencies from JD tuples. Moreover, we adopt a two-stage learning strategy for skill-aware recommendation, in which we utilize the skill distribution to guide JD representation learning in the recall stage, and then combine the user profiles for final prediction in the ranking stage. Consequently, we can embed rich contextual semantic representations for learning JDs, while skill-aware recommendation provides effective JD-user joint representation for click-through rate (CTR) prediction. To validate the superior performance of our method for job recommendation, we present a thorough empirical analysis of large-scale real-world and public datasets to demonstrate its effectiveness and interpretability.
\end{abstract}

}

\maketitle

\IEEEdisplaynontitleabstractindextext

\IEEEpeerreviewmaketitle

\IEEEraisesectionheading{\section{Introduction}\label{sec:introduction}}
Job recommendation aims at providing the right jobs to the right job seekers. In recent years, online recruitment data has experienced explosive growth. According to the report from The Insight Partners~\cite{theinsightpartners}, the global online recruitment market size is expected to grow from \$29.29 billion in 2021 to \$47.31 billion by 2028. As a result, it is crucial for recruitment platforms to develop effective job recommendation systems that not only help companies quickly recruit candidates for specific positions, but also meet the needs of job seekers for an efficient and personalized job search experience.

\begin{figure}[t]
	\centering
	\includegraphics[width=90mm]{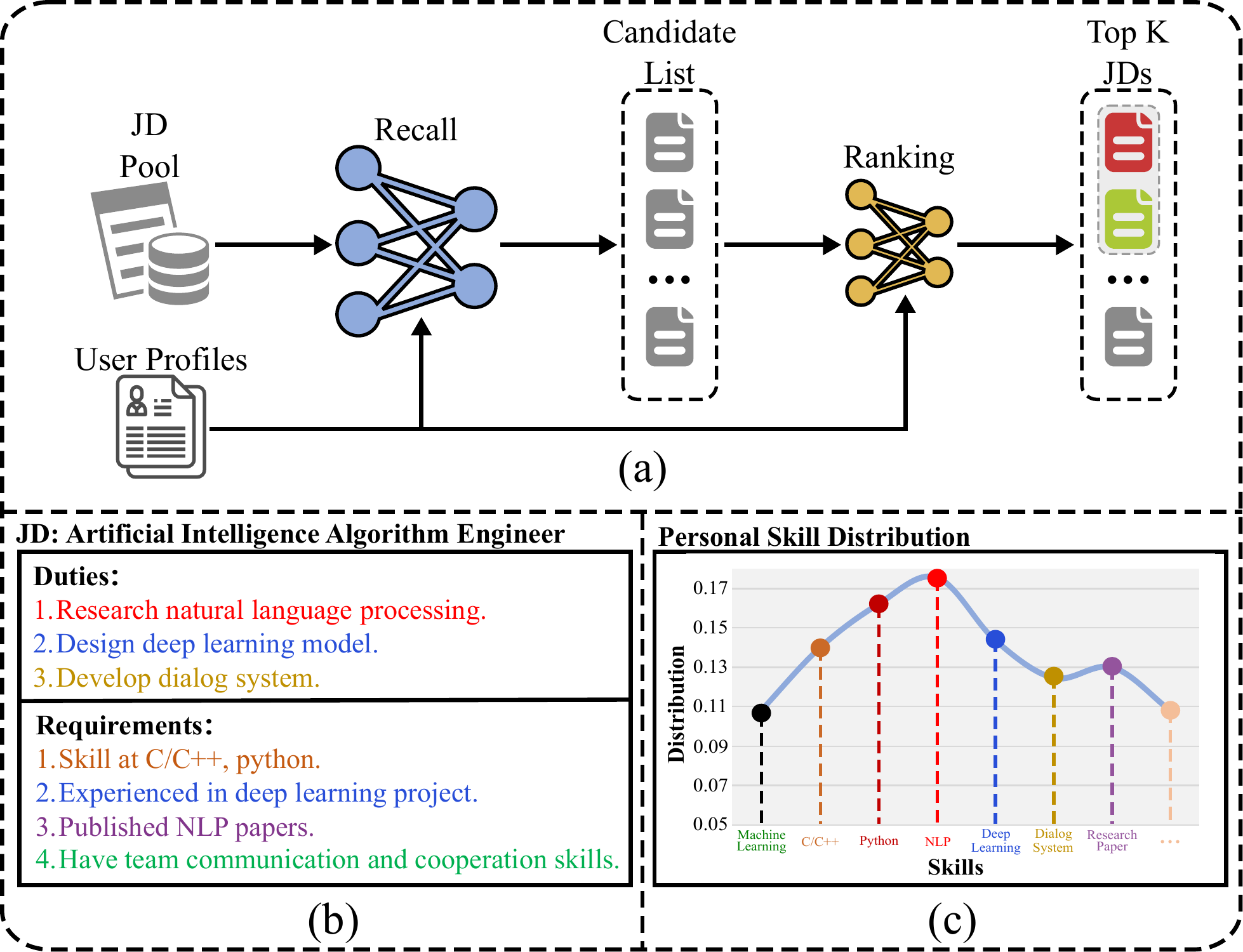}
	\caption{A motivating example of two-stage skill-aware job recommendation (a). The JD contains multiple items including duties and requirements (b), and the corresponding user can be represented with personal skill distribution (c). Note that actually the items in JD correspond to the skill labels, and the degrees of demands correspond to the skill distributions. }\label{fig:f1}
\end{figure}

In real-world recruitment scenarios, the available recruitment records typically only include job descriptions (JDs), user profiles, and click data. It is worth noting that users generally do not upload complete user profiles, primarily because complete user profiles may involve sensitive information, which could lead to privacy concerns or identity theft in case of improper use. Therefore, user profiles are usually summarized as personal skill distribution. With the rapid development of deep learning (DL), intelligent recommendation systems have revolutionized the recruitment field~\cite{bian2019domain,yao2022knowledge,bian2020learning,yao2021interactive}. A natural and straightforward idea is to extract available information from JDs and user profiles, and rely on many existing sophisticated recommendation methods to accomplish job recommendation. For example, content-enriched recommendation approaches can model semantic relevance between user profiles and JDs from two aspects: the general features of user profiles and JDs, as well as the textual content information~\cite{wu2022survey}. The former focuses on the feature interactions of user profile and JD, e.g.,~\cite{cheng2016wide,guo2017deepfm,wang2018tem} explore the possibility of adopting neural models to automatically discover complex higher-order feature interactions for click-through rate (CTR) prediction and recommendation. The latter focuses more on multi-level automatic representation learning of textual content, e.g.,~\cite{QinZXZJCX18,ZhuZXMXDL18} employ the specially designed neural network to model the talent resumes and job descriptions respectively, which are jointly trained as a binary classification problem (i.e., consistent or inconsistent). Considering that these approaches cannot effectively mine users’ interest preferences, a few approaches also explore to generate recommendation lists in a two-stage manner, which contains a recall stage for forming a candidate set and a ranking stage for ranking candidate items based on their relevance to user interests~\cite{wu2021two,qi2021uni}. However, these approaches are problematic since the JD itself is insufficiently informative, and there is a natural heterogeneous gap between JD (represented by short texts in Figure \ref{fig:f1} (b)) and user profiles (represented by skill distributions in Figure \ref{fig:f1} (c), which inevitably leads to a decline in recommendation effectiveness.


To this end, this paper proposes a skill-aware recommendation method with a semantic-enhanced transformer. We first consider enhancing the semantic representation of JD by aggregating complementary information from neighbor JDs, which provides a more comprehensive understanding of the duties and requirements for a specific position, since neighbor JDs typically contain information related to the same domain or position. Specifically, we first model the relative items of each JD, and then adopt an encoder with the local-global attention mechanism to better mine the intra-job and inter-job dependencies from JD tuples. Moreover, in order to better mitigate the heterogeneous gap between JD and user profiles, as well as effectively discover users' interest preferences, our idea is to employ a two-stage learning strategy for skill-aware job recommendation. To be specific, as shown in Figure \ref{fig:f1} (a), we first utilize user profiles (i.e., personal skill distributions) to guide the representation learning of JDs in the recall stage, which enables to recall a set of candidate JDs in the embedding space according to the relevance metric function (i.e., cosine similarity). Then, in the ranking stage, we further predict the click-through rate between candidate JDs and the user for a personalized job recommendation. The major contributions can be summarized as follows:
\begin{itemize}
        \item Develop the semantic-enhanced transformer, which encodes the job descriptions with the designed local-global attention mechanism for rich contextual semantics;
        \item Propose the JobFormer, a two-stage method for job recommendation, which leverages skill-aware JD representation to mitigate the heterogeneous gap between job descriptions and user profiles, as well as to promote recommendation performance;
	\item Empirically show the superiority of JobFormer on real-world datasets. We achieve state-of-the-art results and better interpretability for job recommendation.
\end{itemize}

\section{Related Work}
In this paper, we aim to learn an effective job recommendation method with skill-aware JD representation. Therefore, our study is related to job recommendation systems and JD representation learning.

\subsection{Job Recommendation Systems}
Job recommendation is a core component of recruitment platforms, and it has been extensively studied in the literature~\cite{kenthapadi2017personalized,chou2020based}. Early approaches treated this problem as a job-resume matching problem~\cite{diaby2013toward,lu2013recommender}, and obtained matching capabilities based on the collaborative filtering assumption. However, this approach overemphasizes the interaction between user profiles and job postings, which can result in limited recommendation performance when interaction data is sparse. To mitigate this challenge, recent studies have focused more on utilizing intelligent techniques to mine textual information, aiming to enhance the semantic representation of job and user profiles. Around this problem,~\cite{deshpande2020mitigating} employed TF-IDF statistical approach to encode JDs and resumes.~\cite{zhang2016glmix} developed a generalized linear mixed model (GLMix), a fine-grained model at the user or item level, in the LinkedIn job recommendation system, and generated 20\% to 40\% more job applications for job seekers. Thanks to the advances in deep neural networks (DNNs) that are extensively used in the field of natural language processing (NLP), DNN-based approaches are proposed and have demonstrated state-of-the-art results. For instance,~\cite{ZhuZXMXDL18} developed a novel end-to-end neural model, which projects both job postings and candidate resumes onto a shared latent representation for joint representation learning.~\cite{QinZXZJCX18} designed an ability-aware neural network, which extracts the ability-aware representations for job postings and resumes simultaneously by hierarchical representation structures. These methods emphasize the significance of considering effective representations of multi-modal input (i.e., job descriptions and resumes) in job recommendation tasks. Nevertheless, previous job recommendation approaches usually only consider the relevance between JDs and user profiles, without paying more attention to users’ interest preferences for JDs. Moreover, due to the complex heterogeneity between JDs and user profiles (e.g., personal skill distribution), it remains challenging to improve job recommendation performance.

\subsection{JD Representation with Deep Learning}
Generally, the JD representation problem based on textual data can be categorized as the tasks of text mining, which is highly relevant to Natural Language Processing techniques, such as text classification~\cite{lu2020multi,tian2019hierarchical}, machine translation~\cite{xia2019tied,weng2020acquiring}, and reading comprehension~\cite{berant2014modeling,hermann2015teaching}. Recently, due to the advanced performance and flexibility of deep learning, more and more researchers have attempted to leverage deep neural networks to address text mining problems. In contrast to traditional approaches that heavily rely on effective manually designed representations and input features (e.g., N-gram model~\cite{wang2012baselines}, Bag-of-words model~\cite{lan2008supervised} and parse trees~\cite{cherry2008discriminative}), the deep learning-based approaches can automatically learn effective feature representations from a large-scale text corpus. 

Among various deep learning models, traditional deep learning models (e.g., convolutional neural network (CNN)~\cite{lecun1998gradient} and recurrent neural network (RNN)~\cite{elman1990finding}) and transformer-based models are two representative and extensively used approaches, which can provide practical ways for JD representation from different perspectives. Specifically, CNNs can effectively extract local semantics and hierarchical relationships in textural data. For instance,~\cite{shen2018joint} proposed to encode the job based on CNN.~\cite{ZhuZXMXDL18} have shown that the power of CNN on person-job fit tasks, even only using a few one-dimensional convolutional layers to learn JD representation. Furthermore, RNN-based models have also achieved remarkable performance. For example,~\cite{QinZXZJCX18} designed a word-level semantic representation for both job requirements and job seekers' experiences based on the Recurrent Neural Network. Similarly,~\cite{yan2019interview} adopt the RNNs with GRU units to propagate information along the word sequence of job posting. Compared with traditional deep learning models, transformer-based models are more "natural" in modeling sequential textual data, and learning the contextual dependency and global semantic representation with self-attention mechanism. For example,~\cite{bian2020learning} developed a hierarchical self-attention text representation model for developing the semantic matching model, in which a BERT-based encoder is first adopted to represent jobs, and then a transformer-based encoder is used to represent the overall text document based on learned sentence embeddings. Although transformer processes textual data efficiently, the information deficit of JD itself reduces the parsing of JD semantics.

In this paper, we follow some outstanding ideas in the above works according to the properties of job recommendation and propose a two-stage method JobFormer based on the semantic-enhanced transformer with skill-aware JD representation. Therefore, JobFormer can not only improve the performance of job recommendation, but also enhance the model interpretability in practical scenarios.


\begin{figure*}[t]
	\centering
	\includegraphics[width=180mm]{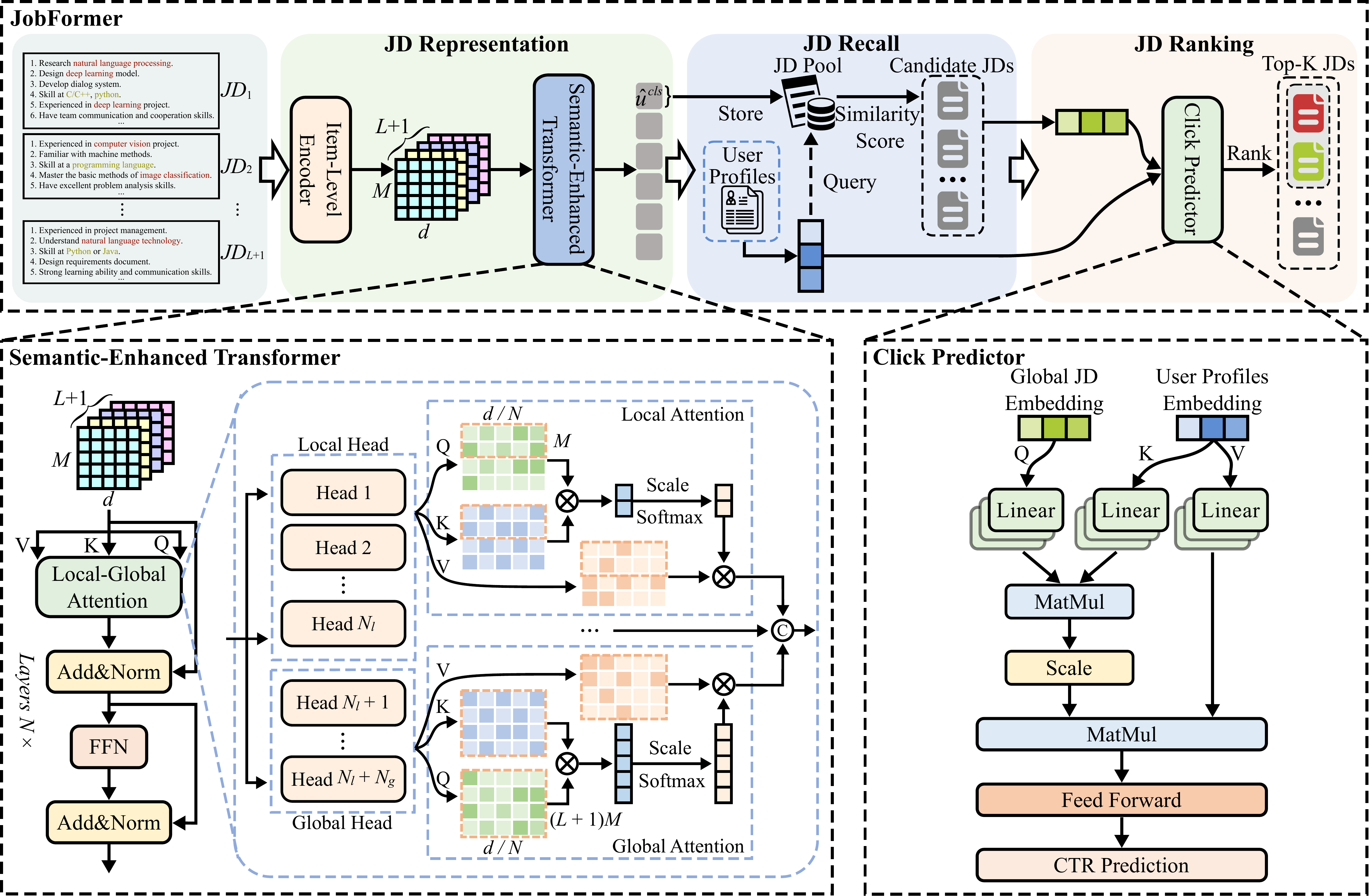}
	\caption{An illustration of the proposed JobFormer, which includes a recall stage for candidate JDs generation and a ranking stage for candidate JDs ranking. In the recall stage, the JD and its neighbors constitute the JD tuple, and the item-level encoder aims for the item representation, which acts as the input token for the semantic-enhanced transformer. Then the designed semantic-enhanced transformer encodes both the intra-job and inter-job information to acquire more discriminative JD representation, which is further recalled as candidate JDs according to the JD-user cosine similarity. Lastly, in the ranking stage, recalled candidate JDs are combined with user profiles for CTR prediction via a click predictor and ranked for a personalized job recommendation.}\label{fig:f2}
\end{figure*}

\section{Preliminaries}
In company talent management, the available JD-user records usually include job descriptions, user profiles (i.e., personal skill distribution) and click data. Specifically, a job description contains multiple items with short text forms, which describe duties and requirements. Without any loss of generality, we utilize $\jj = \{\jj^1,\jj^2,\cdots,\jj^M\}$ to denote the items (e.g., the JD about artificial intelligence algorithm engineer as shown in Figure \ref{fig:f1} (b)), $M$ denotes the total number of duties and requirements, and we fix $M$ as the maximum number of items for all JDs with padding mask as~\cite{VaswaniSPUJGKP17}. For user profiles, we adopt all the skills summarized as global label space (i.e., $C$ skills totally), and then acquire the skill ratings from experts. Lastly, we normalize the summarized skill ratings as the label distribution with softmax operator according to~\cite{GengJ13}, i.e., $\y = \{y^1,y^2,\cdots,y^C\}$, satisfying the constraint $y^c \in [0,1]$ and $\sum_c y^c = 1$. To simplify our problem, we assume that a job can be represented by its duties and requirements, and the skill distribution of a candidate can mainly reflect his competency. 

As a matter of fact, as shown in Figure \ref{fig:f1} (b), the duties and requirements in JD correspond to the personal skills, e.g., "Design deep learning model" indicates that the user needs to have "deep learning" skill. The degree of demands corresponds to the skill distributions, e.g., considering that "deep learning" is repeatedly mentioned in the JD, and the demand is high (i.e., "Design deep learning model" in duties and "Experienced in deep learning project" in requirements), the degree of skill "deep learning" should be high. A straightforward approach is to utilize label distribution learning (LDL) to predict the skill distribution of the input JDs and match it with the user profiles (i.e., personal skill distribution) according to the cosine similarity. However, this approach only recommends JDs to users from the perspective of relevance, which lacks the mining of users’ interest preferences. Therefore, our idea is first to recall a set of candidate JDs and then utilize the click data to optimize the ranking ability of the model on the candidate JDs for a personalized job recommendation. Along this line, we can formally define the problems:


{\defn(Skill-Aware Representation Learning for JD Recall). 
	{Given a set of successful person-job records $\D$, each record $(\jj,\y) \in \D$ is the corresponding JD and skill distribution. The target of JD recall can be formulated as learning a predictive model $f$ for predicting the skill distribution of the input $\jj$, and then recalling a set of candidate JDs from a large-scale JD pool based on their relevance to user profiles.}}

{\defn(Click-Through Rate Prediction for JD Ranking). 
	{Given a set of candidate sets $\D'$ with their corresponding click data, $\D'\subset\D$. JD ranking aims to further discover user-interesting JDs from candidate JDs, which can be defined as a CTR prediction task, i.e., predicting the probability of a user clicking on the candidate JDs.}}

\section{JobFormer}
As shown in Figure \ref{fig:f2}, following a widely used paradigm in real-world recommendation systems, JobFormer contains a recall stage for candidate JDs generation and a ranking stage for candidate JDs ranking. Specifically, in the recall stage, we first leverage the TextCNN~\cite{KalchbrennerGB14} to encode the JD for diverse item-level representations, which are fed into the local-global transformer to capture rich contextual semantics. The learned JD representations are further calculated similarity scores with user profiles (e.g., personal skill distribution) to recall candidate JDs from a large-scale JD pool. Finally, in the ranking stage, candidate JDs are combined with user profiles for CTR prediction with a click predictor. Next, we will describe each component of our JobFormer in detail.

\subsection{Item-Level Encoder}
As shown in Figure \ref{fig:f1} (b), each job description $\jj$ includes a set of items, including the duties and requirements. It is notable that these items are in short text forms and have no contextual information, which may lead to semantic confusion if we directly concatenate these items into long sentence for representation learning. For example, the word "Design" in "Design deep learning model" only describes the duty of deep learning model, and has no contribution to other duties or requirements, so there is no contextual relationship. Based on this idea, we need to model the items separately to obtain the item-level representations.

Without any loss of generality, we adopt a shared TextCNN to process the items separately. In detail, given an item $\jj^m$ with $S$ words, the corresponding matrix can be represented as: $\jj^m = [\x_1 , \x_2, \cdots, \x_S] \in \R^{S \times d}$. Then we apply two one-dimensional convolutional layers on the input layer considering that one-dimensional convolution can deal with an unfixed-length sequence~\cite{KalchbrennerGB14}. To reduce the training cost, we apply Batch Normalization~\cite{IoffeS15} followed by a Rectified Linear Unit (ReLU) layer~\cite{NairH10} and a one-dimensional max-pooling layer on the outputs of one-dimensional convolutional layers. Therefore, each item is inputted to the shared TextCNN:
\begin{equation}\label{eq:e}
\begin{split}
\uu^m = TextCNN(\jj^m) \nonumber
\end{split}
\end{equation}
where $\uu^m \in \R^d$ denotes the representations of $m-$th item.

\subsection{Semantic-Enhanced Transformer} 
Based on the item-level representations, the key challenges to encode the JD are: 1) Various importance. Different duties and requirements have various importance in job description. Take the JD in Figure \ref{fig:f1} (b) as an example, "Design deep learning model" and "Experienced in deep learning project" are more important than "Priority for published papers" considering that deep learning skill is repeatedly mentioned and has a high demand (e.g., "Design" and "Experienced"), while "published papers" is a supplementary condition. 2) Deficient information. Single JD may ignore some derived information. Take the JD in Figure \ref{fig:f1} (b) as an example, many items are related to "machine learning" skill, which is needed for the post (i.e., the successfully accepted user is considered for this skill). To overcome these problems, we design the transformer with local-global attention heads to measure the importance of intra-job items and integrate the inter-job information. 

\textbf{Local Encoder.} To comprehensively encode each JD by considering the dependencies between items, we employ the transformer encoder~\cite{VaswaniSPUJGKP17} as the backbone, which can encode the relationships among items by adopting the self-attention mechanism. Specifically, as shown in Figure \ref{fig:f2}, with the item representations, a job can be denoted as $U_l = [\uu^1, \uu^2, \cdots, \uu^M] \in \R^{M \times d}$, where $d$ is the dimension of hidden states. The identical block contains two sub-layers: 1) The first sub-layer utilizes multi-head attention to learn the correlated representations. 2) The second sub-layer adopts position-wise feed-forward network (FFN). In multi-head attention layer, the input representations can be used to compute three matrices: $Q$, $K$, and $V$ corresponding to queries, keys, and values. The dot-product similarity between queries and keys determines attention distributions:
\begin{equation}\label{eq:e0}\small
\begin{split}
Q_l & = U_l W_{Q_l}, \quad K_l = U_l W_{K_l}, \quad V_l = U_l W_{V_l}, \\
A_l & = \frac{Q_l K_l^\top}{\sqrt{d_{N_l}}} \quad Att(U_l) = \sigma(A_l)V_l, \\
\end{split}
\end{equation}
where $Q_l \in \R^{M \times d_{N_l}}$, $K_l \in \R^{M \times d_{N_l}}$, $V_l \in \R^{M \times d_{N_l}}$, and $W_{Q_l} \in \R^{d \times d_{N_l}}, W_{K_l} \in \R^{d \times d_{N_l}}, W_{V_l} \in \R^{d \times d_{N_l}}$ are learnable matrices. $N_l$ denotes the number of local heads. The activation function $\sigma$ can be used as softmax here. It is notable that the multi-head attention is defined as the local attention here in literature, which aims to encode the intra-job information.


\textbf{Global Encoder.} To introduce the extra neighbor JDs as complementary information, we further propose the joint modeling strategy with local-global attention. In detail, we first select $L$ neighbors for $\jj$, i.e., $\N(\jj) = \{\jj_{1},\jj_{2},\cdots,\jj_{L}\}$. Without any loss of generality, we adopt the euclidean distance according to the global embedding $\hat{\uu}^{cls}$ and title of JD (e.g., the "Artificial Intelligence Algorithm Engineer" as shown in Figure \ref{fig:f1} (b)), i.e., $\|{\hat{\uu}}^{n_1,cls} - {\hat{\uu}}^{n_2,cls}\|_2^2, \quad t(\uu^{n_1}) = t(\uu^{n_2})$, where $t(\cdot)$ represents the title representation with one-hot form, $n_1$ and $n_2$ represent the index of $\hat{\uu}$. Thereby, we can concatenate $L+1$ JDs as input with special token $[SEP]$ for separation, i.e., $U_g = [U_l, \N(U_l)]^\top \in \R^{(L+1)M \times d}$. As shown in Figure \ref{fig:f2}, we add $N_g$ parallel heads as global attention. The dot-product similarity can be reformulated as:
\begin{equation}\label{eq:e7}\small
\begin{split}
Q_g & = U_g W_{Q_g}, \quad K_g = U_g W_{K_g}, \quad V_g = U_g W_{V_g}, \\
A_g & = \frac{Q_g K_g^\top}{\sqrt{d_{N_g}}} \quad Att(U_g) = \sigma(A_g)V_g, \\
\end{split}
\end{equation}
where $Q_g \in \R^{(L+1)M \times d_{N_g}}$, $K_g \in \R^{(L+1)M \times d_{N_g}}$, $V_g \in \R^{(L+1)M \times d_{N_g}}$, and $W_{Q_g} \in \R^{d \times d_{N_g}}, W_{K_g} \in \R^{d \times d_{N_g}}, W_{V_g} \in \R^{d \times d_{N_g}}$ are learnable matrices. The activation function $\sigma$ can be used as softmax here. Finally, local-global attention is composed of $N = N_l + N_g$ parallel heads, and $d_{N_l} = d_{N_g} = d/N$. For each JD, the results of local head and corresponding global head are concatenated, and the FFN can be reformulated as:
\begin{equation}\label{eq:e2}\small
\begin{split}
MultiAtt(U) & =  [Att(U_l)_1,\cdots, Att(U_l)_{N_l},\\ & Att(U_g)_1[id(U_l)],\cdots,Att(U_g)_{N_g}[id(U_l)]] W_U \\
FFN(MultiAtt(U)) & =  \max(0, MultiAtt(U) W_1 + b_1)W_2 + b_2,
\end{split}
\end{equation} 
where $id(U_l)$ denotes the corresponding index of local $U_l$ in $Att(U_g)$. $W_1$, $W_2$, and $W_U$ are matrices for linear transformation, $b_1$ and $b_2$ are the bias terms. Meanwhile, each sub-layer is followed by dropout~\cite{SrivastavaHKSS14}, shortcut connection~\cite{HeZRS16}, and layer normalization~\cite{BaKH16}. We can also add the special token $[CLS]$ for each JD to learn the global representations, and the attention value depicts the importance of each item. Finally, for each JD, we can acquire the global JD embedding, i.e., $\hat{\uu}^{cls}$, from the $[CLS]$ token, and individual embedding from other tokens. 

Therefore, the local head attentions are responsible for capturing local dependencies based on local details, i.e., the intra-job items, and global head attentions are designed to model the long-term dependencies between JDs, i.e., the inter-job items. The combination of local and global attention enables our JobFormer to dynamically model local items and capture the global dependencies of similar JDs. Consequently, for each JD $\jj$, we can acquire both the JD-level and item-level representations, i.e., $\hat{\jj} = [\hat{\uu}^{cls}, \hat{\uu}^1, \hat{\uu}^2, \cdots, \hat{\uu}^M] \in \R^{(M+1) \times d}$, $\hat{\uu}^{cls} \in \R^d$ is the JD global representation.

\subsection{Skill-Aware Representation Learning for JD Recall}
With the JD-level representations $\hat{\uu}^{cls}$, we can directly predict the skill distribution with a simple classifier, e.g., a fully connected network $g$ with softmax operator. For simplicity, we can adopt the KL-divergence~\cite{mackay2003information} between ground-truth and prediction:
\begin{equation}\label{eq:e3}\small
\begin{split}
\ell = KL(\y,g(\hat{\uu}^{cls})) = \sum_c \y^c \log(\frac{\y^c}{g^c(\hat{\uu}^{cls})})
\end{split}
\end{equation}
$KL(a,b) = a \log \frac{a}{b}$ is the KL-divergence that penalizes difference. However, the direct prediction ignores the correlations between different skills, i.e., a skill can help to learn another skill under certain conditions. For example, as shown in Figure \ref{fig:f1}, when the user has a high description degrees on skills "Deep Learning" and "Dialog System", the skill "Natural Language Processing" is more likely to have a higher description degree than skill "Software Engineering". Because "Deep Learning" and "Dialog System" are usually related to "Natural Language Processing". Therefore, taking skill correlations into consideration can include more data information and achieve better performance~\cite{JiaLLZ18,YangFZLJ21,HuangZ12}, we consider two aspects in the recall stage: 1) skill correlation enhancement. 2) relation consistency.

\textbf{Skill correlation enhancement.} In most real-world applications, skill correlations are usually local, i.e., different instances have specific skill correlations. For example, the needed users who are proficient in natural language processing should have higher similarity values among “Deep Learning”, “Dialog System” and “Natural Language Processing” than the needed users who are only familiar with natural language processing. Based on this idea, we collect additional information about the accepted users to assist in predicting the skill distribution, including the position name (which can reflect the skills the user has mastered) and position level (which can reflect the user's skill proficiency). In detail, we first encode this additional user information ${\e}^{aui}$ with a fully connected layer and then adopt a bidirectional ranking loss~\cite{faghri2017vse++} with margin $\alpha$ to match JD (i.e., $\hat{\uu}^{cls}$) and additional user information (i.e., ${\e}^{aui}$):
\begin{equation}\label{eq:e12}
\begin{split}
M =  [\alpha-s({\e}^{aui},\hat{\uu}^{cls})+s({\e}^{aui},{\hat{\uu}^{cls}}_{*}]_+ \\
+ [\alpha-s({\e}^{aui},\hat{\uu}^{cls})+s({\e}^{aui}_{*},\hat{\uu}^{cls})]_+
\end{split}
\end{equation}
where $[x]_+ \equiv max(x,0)$ and $s$ is a similarity score function (i.e., cosine similarity). ${\hat{\uu}^{cls}}_{*}$ and ${\e}^{aui}_{*}$ represent the corresponding hardest negative JD and hardest negative additional user information within a minibatch, respectively.

\textbf{Relation consistency.} To constrain the structure of neighbors, we define a relation consistency constraint using a metric learning-based constraint, inspired by the linguistic structuralism~\cite{Matthews2001A} that relations can better present the knowledge than individual examples. Specifically, each JD and its neighbors can be denoted as a bag of $L+1$ instances, i.e., $\Psi(\jj)$, and the pairwise similarity of JDs representations and predictions should be consistent. Therefore, we constrain the KL divergence of the similarity vectors calculated by the predictions and representations. The JD representations can be denoted as $\hat{\uu}^{i,cls}$ and the predictions can be formulated as $g(\hat{\uu}^{i,cls})$, where $ \hat{\uu}^{i,cls} = SE-Transformer(TextCNN(\jj_i))$, $SE-Transformer$ denotes the semantic-enhanced transformer. Therefore, the objective of relation consistency can be formulated as:
\begin{equation}\label{eq:e6}\small
\begin{split}
R = &  KL(\Phi(g(\hat{\uu}^{1,cls}), \cdots,g(\hat{\uu}^{L+1,cls})),  \Phi(\hat{\uu}^{1,cls},\cdots,\hat{\uu}^{L+1,cls})),  
\end{split}
\end{equation}
$\Phi$ is a relation prediction function with softmax operator, which measures the relation energy of the given tuple. In detail, $\Phi$ aims to measure the similarities, using the predictions as an example:
\begin{equation}\small
\begin{split}
\Phi(g(\hat{\uu}^{1,cls}), \cdots,g(\hat{\uu}^{L+1,cls})) &= [q_{{n_1,n_2}}] \quad {n_1,n_2 \in [1,\cdots,L+1]} \\
q_{{n_1,n_2}} &= \frac{exp(d_{{n_1,n_2}})}{\sum exp(d_{\cdot})} \\ \nonumber
\end{split}
\end{equation}
where $d_{{n_1,n_2}} = KL(g(\hat{\uu}^{n_1,cls}),g(\hat{\uu}^{n_2,cls}))$ measures the distance. $q_{{n_1,n_2}}$ denotes the relative instance-wise similarity. Finally, we pull the $[q_{{n_1,n_2}}]$ into vector form. $\Phi(\hat{\uu}^{1,cls},\cdots,\hat{\uu}^{L+1,cls})$ is calculated in the same way, with $d_{{n_1,n_2}} =  \|\hat{\uu}^{n_1,cls} - \hat{\uu}^{n_2,cls} \|_2$. Since the structure has higher-order properties than a single output, it can transfer knowledge more effectively and is more suitable for consistency measures. We define the total loss by combining the Eq. \ref{eq:e3}, Eq. \ref{eq:e12}, and Eq. \ref{eq:e6}:
\begin{equation}\label{eq:e8}
\begin{split}
L =  \sum_{\jj} \ell(\jj) +\lambda M(\jj) + \mu R(\jj).
\end{split}
\end{equation}
where $\lambda$ and $\mu$ is the balance parameter. To effectively minimize the target function with the guidance of auxiliary neighbors, each training batch consists groups of JDs. Each group contains a central JD together with its neighbor JDs as~\cite{JiaLZLH21}. During inference, each test JD can be conditioned by a set of neighbors from training JDs to construct JD tuples. Finally, we compare the relevance degree between predicted skill distribution and personal ground-truth, i.e., $sc(g(\hat{\uu}^{cls}),\y)$, $sc$ denotes the cosine similarity function, and then recall candidate JDs from a large-scale JD pool according to the similarities.

\subsection{Click-Through Rate Prediction for JD Ranking}
The ranking stage aims to further discover user-interesting JDs from a small number of candidate JDs. As shown in Figure \ref{fig:f2}, the candidate JD and user profiles (i.e., personal skill distribution) are used to compute a click score via click predictor for personalized JD ranking. More specifically, we first train a cross-attention module to obtain joint embedding using candidate JD global embedding $\hat{\uu}^{cls} \in \R^d$ and user profiles $\s \in \R^n$, where $d$ and $n$ represent the dimension of embedding and the total number of skills, respectively:
\begin{equation}\label{eq:e9}\small
\begin{split}
Q_c & = (\hat{\uu}^{cls})^\top W_{Q_c}, \quad K_c = \s^\top W_{K_c}, \quad V_c = \s^\top W_{V_c}, \\
A_c & = {Q_c^\top K_c}, {\e}^{joint} = \sigma(A_c)V_c^\top,   \\
\end{split}
\end{equation}
where $Q_c \in \R^{1 \times d}$, $K_c \in \R^{1 \times n}$, $V_c \in \R^{1 \times n}$, and $W_{Q_c} \in \R^{d \times d}, W_{K_c} \in \R^{n \times n}, W_{V_c} \in \R^{n \times n}$ are learnable matrices. The activation function $\sigma$ can be used as softmax here. Then, the joint embedding ${\e}^{joint} \in \R^d$ is fed into two fully connected layers with sigmoid function to output the predicted click probability:
\begin{equation}\label{eq:e10}\small
\begin{split}
\hat{y}_{click} = sigmoid(FC({\e}^{joint}))
\end{split}
\end{equation}
Furthermore, we define the click-through rate prediction task as a binary classification problem, where a JD-user click interaction is assigned a target value 1, otherwise 0. Specifically, we use the cross-entropy as the loss function:
\begin{equation}\label{eq:e11}\small
\begin{split}
L = -(\sum_{(jd,user) \in \R^+} log(\hat{y}_{click}) + \sum_{(jd,user) \in \R^-} log(1-\hat{y}_{click}))
\end{split}
\end{equation}
where $\R^+$ and $\R^-$ are the positive and negative click records. Finally, the click scores of candidate JDs are used for personalized ranking.
\section{Experiments}
In this section, we demonstrate the effectiveness of JobFormer by verifying the following problems: 
\begin{itemize}
	\item The recall and ranking performance compared with state-of-the-art baselines;
	\item The performance compared with various variants;
	\item Sensitivity analyses of parameters;
	\item Interpretability of JobFormer.	
\end{itemize}

\begin{table*}[t]{\small
		\centering
        \renewcommand{\arraystretch}{1.3}
		\caption{Experimental results of different approaches on JD recall task. Higher Recall and NDCG rates mean better performance.}
		\label{tab:tab1}
		\begin{tabular*}{1\textwidth}{@{\extracolsep{\fill}}@{}c|@{\hspace{0.05cm}}@{}c@{\hspace{0.05cm}}|@{}c@{\hspace{0.05cm}}|@{}c@{\hspace{0.05cm}}|@{}c@{\hspace{0.05cm}}|@{}c@{\hspace{0.05cm}}|@{}c@{\hspace{0.05cm}}|@{}c@{\hspace{0.05cm}}|@{}c@{\hspace{0.05cm}}|@{}c@{\hspace{0.05cm}}|@{}c@{\hspace{0.05cm}}}
            \Xhline{1pt}
			\multirow{2}{*}{Methods} & \multicolumn{10}{c}{Metric} \\
            \Xcline{2-11}{0.1pt}
			& Recall@20 & Recall@40 & Recall@60 & Recall@80 & Recall@100 & NDCG@20 & NDCG@40 & NDCG@60 & NDCG@80 & NDCG@100\\
            \Xhline{0.5pt}
			NPA &61.76 &77.68 &81.78 &87.39 &92.39 &28.18 &31.46 &32.18 &33.09 &33.86 \\
			NAML &62.46 &78.68 &87.19 &91.89 &95.20 &32.29 &36.33 &37.96 &38.02 &38.72 \\
			NRMS &67.67 &81.18 &86.69 &88.79 &93.09 &33.12 &35.45 &37.88 &38.33 &38.72 \\
                CNE-SUE &61.66 &76.68 &87.99 &93.99 &96.60 &27.70 &30.53 &31.50 &31.84 &32.50 \\
            \Xhline{0.5pt}
			UNBERT &51.43 &62.14 &69.34 &76.95 &85.36 &23.77 &27.66 &29.71 &30.76 &31.27 \\ 
			MINER &59.36 &76.08 &85.29 &92.89 &96.60 &27.94 &31.34 &32.97 &34.21 &34.78 \\
            \Xhline{0.5pt}
			LSMT+ &62.11 &79.45 &86.71 &91.05 &93.02 &29.17 &31.51 &33.11 &34.28 &34.81 \\
			TextCNN+ &67.74 &83.36 &90.12 &93.95 &96.11 &32.44 &35.69 &36.94 &37.57 &37.90 \\
			local-global+  &68.85 &83.66 &90.62 &94.95 &96.98 &33.06 &35.92 &37.56 &37.92 &38.29 \\
                w/o $M$ &69.45 &83.26 &90.52 &94.65 &97.17 &33.21 &36.81 &38.11 &38.52 &39.12  \\
			w/o $R$ &69.65 &82.86 &90.52 &\bf95.16 &97.27 &34.10 &36.94 &38.14 &38.70 &39.17  \\
            \Xhline{0.5pt}
			JobFormer &\bf69.98 &\bf83.98 &\bf90.97 &95.01 &\bf97.29 &\bf34.30 &\bf37.21 &\bf38.42 &\bf39.14 &\bf39.47 \\	 		
            \Xhline{1pt}
	\end{tabular*}}
\end{table*}  

\begin{table}[t]
        \small
	\centering
        \renewcommand{\arraystretch}{1.3}
		\caption{Experimental results of different approaches on JD ranking task. Higher AUC and MRR mean better performance.}
		\label{tab:tab2}
		\begin{tabular}{c|c|c}
            \Xhline{1pt}
			\multirow{2}{*}{Methods} & \multicolumn{2}{c}{Metric} \\
            \Xcline{2-3}{0.1pt}
			& AUC & MRR\\
            \Xhline{0.5pt}
                NPA &75.52 &28.53 \\
                NAML &75.48 &28.39 \\
                NRMS &76.88 &29.11 \\
                CNE-SUE &74.87 &28.06 \\
            \Xhline{0.5pt}
                UNBERT &68.16 &19.45 \\
                MINER &73.21 &26.33 \\
            \Xhline{0.5pt}
                FRNet &76.54 &29.18 \\
                MaskNet &74.93 &27.92 \\
            \Xhline{0.5pt}
            JobFormer &\bf77.58 &\bf29.82 \\
            \Xhline{1pt}
	\end{tabular}
\end{table} 

\begin{table*}[t]{\small
		\centering
        \renewcommand{\arraystretch}{1.3}
		\caption{Experimental results of different local-global heads and neighbors on JD recall task. (JF: JobFormer)}
		\label{tab:tab3}
		\begin{tabular*}{1\textwidth}{@{\extracolsep{\fill}}@{}c|@{\hspace{0.05cm}}@{}c@{\hspace{0.05cm}}|@{}c@{\hspace{0.05cm}}|@{}c@{\hspace{0.05cm}}|@{}c@{\hspace{0.05cm}}|@{}c@{\hspace{0.05cm}}|@{}c@{\hspace{0.05cm}}|@{}c@{\hspace{0.05cm}}|@{}c@{\hspace{0.05cm}}|@{}c@{\hspace{0.05cm}}|@{}c@{\hspace{0.05cm}}}
            \Xhline{1pt}
			\multirow{2}{*}{Methods} & \multicolumn{10}{c}{Metric} \\
            \Xcline{2-11}{0.1pt}
			& Recall@20 & Recall@40 & Recall@60 & Recall@80 & Recall@100 & NDCG@20 & NDCG@40 & NDCG@60 & NDCG@80 & NDCG@100\\
            \Xhline{0.5pt}
	        JF ($N_l=4$, $N_g=4$) &68.34 &83.16 &90.02 &94.45 &96.47 &33.60 &36.50 &37.64 &38.36 &38.59 \\	
			JF ($N_l=5$, $N_g=3$) &69.15 &\bf84.03 &90.32 &94.75 &\bf97.31 &33.83 &36.93 &37.94 &38.73 &39.05  \\	
			JF ($N_l=6$, $N_g=2$) &\bf69.98 &83.98 &\bf90.97 &\bf95.01 &97.29 &\bf34.30 &\bf37.21 &\bf38.42 &\bf39.14 &\bf39.49  \\	
			JF ($N_l=7$, $N_g=1$) &68.85 &83.77 &90.24 &94.64 &97.09 &34.04 &36.92 &38.03 &38.79 &39.07\\
            \Xhline{0.5pt}
                JF ($L = 1$) &69.25 &83.26 &90.02 &94.15 &96.27 &33.02 &36.07 &37.26 &37.89 &38.29 \\
			JF ($L = 2$) &\bf69.98 &83.98 &\bf90.97 &95.01 &\bf97.29 &\bf34.30 &\bf37.21 &\bf38.42 &\bf39.14 &\bf39.49 \\
			JF ($L = 3$) &69.95 &\bf84.23 &90.52 &\bf95.03 &97.07 &33.65 &36.62 &38.22 &38.28 &39.03\\
			JF ($L = 4$) &68.75 &83.56 &90.51 &94.45 &96.87 &32.89 &35.97 &37.22 &37.73 &38.19 \\
			JF ($L = 5$) &67.84 &83.16 &90.22 &94.05 &96.91 &32.27 &35.42 &36.56 &37.22 &37.70 \\		
            \Xhline{1pt}
	\end{tabular*}}
\end{table*}

\subsection{Data Description}
We conduct our validation on a real-world talent recruitment dataset, which is provided by a high-tech company in China. To protect the privacy of candidates, all the application records are anonymized by deleting personal information. The dataset contains 37540 successful job applications. Indeed, the low acceptance ($\approx1\%$) clearly validates the importance of job recommendation in online recruitment. In detail, we analyze some statistics of our dataset. We find that the number of applications is relatively steady. Besides, it is notable that each job posting may accept multiple users. We find that the number of job postings with respect to the number of their successfully accepted users, roughly follows a long-tail distribution, and the vast majority of acceptances are controlled within 3 users. Thereby, we randomly select a user to constitute the person-job pair considering that the skill distributions of these users are similar. Finally, 15046 person-job pairs are kept in total.

\subsection{Implementations}

The settings of our experiments include the word embedding of item encoder, structure of semantic-enhanced transformer, and training details for the recall and ranking stages. For item encoder, we first employ TextCNN with two one-dimensional convolutions (i.e., kernel sizes are 2 and 3) as the item-level encoder. The dimension of word is $d = 512$. Note that TextCNN can process an unfixed-length sequence~\cite{KalchbrennerGB14}. The semantic-enhanced transformer is with 4 layers and 8 heads, i.e., $N = 8$, including local heads $N_l =6$ and global heads $N_g = 2$. The classifier $g$ is with two fully connected layers. For semantic-enhanced transformer, we set the maximum number of items in each JD as $M = 40$, the excessive parts are removed. The number of neighbor is $L = 2$, the parameter $\lambda = 0.2, \mu = 0.4$. Following~\cite{GlorotB10}, we initialized all the parameters in JobFormer with uniform distribution in $[-\sqrt{6/(n_{in}+n_{out})},\sqrt{6/(n_{in}+n_{out})}]$, where $n_{in}$, $n_{out}$ denote the number of input and output units, respectively. The number of negative samples associated with positive one is 200 and 3 for the recall and ranking stages, respectively. In all experiments, the batch size is set to 32. The optimization method is Adaptive Moment Estimation (Adam), the learning rate is searched in {0.5, 0.1, 0.05, 0.01, 0.005, 0.001} to find the best settings for each task and annealed by 0.8 every 3 epochs. Finally, we set the learning rate as 0.001. The ratio of dropout is 0.1 and the maximal number of epochs is 30. We run the following experiments with the implementation of an environment on NVIDIA Tesla V100 SXM2 GPUs. The code is available at \url{https://github.com/data-ming-and-application/JAT}.



\subsection{Baseline Approaches}
To verify the effectiveness of JobFormer, we compare it with various baseline methods: 1) Traditional word embedding-based recommendation methods, i.e., NPA~\cite{wu2019npa}, NAML~\cite{wu2019neural1}, NRMS~\cite{wu2019neural2}, and CNE-SUE~\cite{mao2021neural}. 2) BERT-based pre-training recommendation methods, i.e., UNBERT~\cite{zhang2021unbert} and MINER~\cite{li2022miner}. 3) CTR ranking methods, i.e., MaskNet~\cite{wang2021masknet} and FRNet~\cite{wang2022enhancing}:
\begin{itemize}
        \item \textbf{NPA:} a news recommendation method which adopts the personalized attention mechanism to model text semantic information;
        \item \textbf{NAML:} a news recommendation method with attentive multi-view learning to obtain news representation and attentive pooling to learn user representation;
        \item \textbf{NRMS:} a news recommendation method which utilizes multi-head self-attention networks to extract fine-grained representations from the news title and user history respectively;
        \item \textbf{CNE-SUE:} a news recommendation framework consisting of collaborative news encoding and structural user encoding to enhance news and user representation learning;
        \item \textbf{UNBERT:} a BERT-based approach which leverages the pre-trained model to enhance textual representation and capture multi-grained signals at both word-level and news-level;
        \item \textbf{MINER:} a BRET-based pre-training model which employs a poly attention scheme to learn multiple interest vectors for each user;
        \item \textbf{MaskNet:} a CTR ranking framework which takes advantage of instance-guided mask to solve the inefficiency of the feedforward neural network in CTR prediction;
        \item \textbf{FRNet:} a CTR perdition model with a feature refinement module to learn context-aware feature representations by integrating the original and complementary feature representations with bit-level weights.
\end{itemize}
Given the person-job data, the comparison methods for the recall stage consider the JD as input and personal skill distribution as ground-truth for training. For the ranking stage (i.e., CTR prediction task), we utilize the candidate JDs along with the click data to predict the probability of positive feedback, i.e. click, taking place on a JD. 

To measure the performances, we consider two aspects of evaluations: 1) JD recall, which aims to verify that the personal skill distribution can find the candidate JDs from a large-scale JD pool. Following~\cite{KricheneR21}, we adopt the widely-used metrics to evaluate JD recall performance, i.e., Recall@$K$ and NDCG@$K$ (we set $K = 20, 40, 60, 80, 100$). 2) JD ranking, which aims to discover user-interesting JDs from candidate JD pool according to click scores. Specifically, we adopt AUC and MRR to evaluate JD ranking performance. Generally, the larger the values are, the better acquiring results.

\begin{figure*}[t]
	\centering
	\includegraphics[width=180mm]{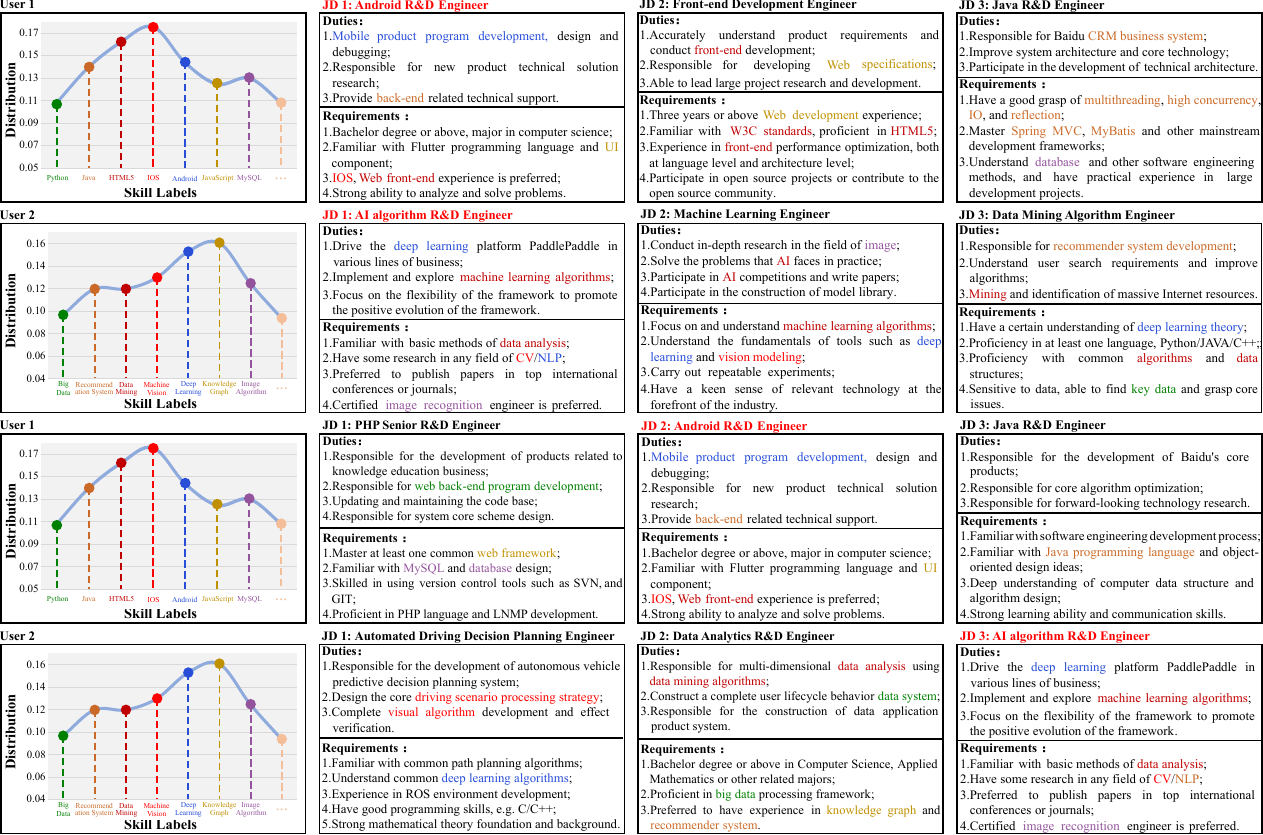}
	\caption{(Best viewed in color when zoomed in.) Qualitative success results of JD recall given user queries. For each user query, we show the top-3 ranked JD text. The first two rows exhibit the results of JobFormer, and the last two rows give the results of the state-of-the-art NRMS model. We observe that our JobFormer can find the correct results (i.e., red marked) in the first-ranked JDs, and NRMS is inferior to JobFormer.}\label{fig:f7}
\end{figure*}


\begin{figure*}[!htb]
	\centering
        \includegraphics[width=180mm]{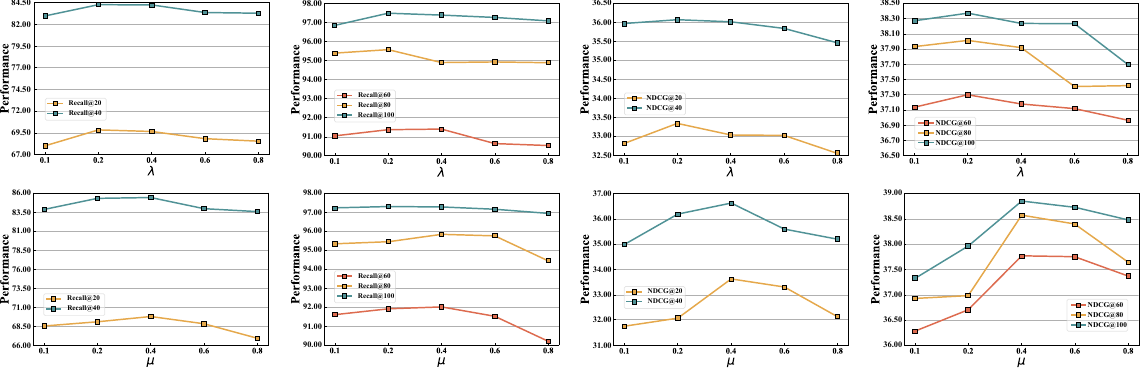}
	\caption{Influence of Balance Parameters. The figures in the first row are the results of $\lambda$, and the second row gives the results of $\mu$.}\label{fig:f3}
\end{figure*}

\begin{table*}[t]{\small
		\centering
        \renewcommand{\arraystretch}{1.3}
		\caption{Experimental results of different approaches on public dataset.}
		\label{tab:tab4}
		\begin{tabular*}{1\textwidth}{@{\extracolsep{\fill}}@{}c|@{\hspace{0.05cm}}@{}c@{\hspace{0.05cm}}|@{}c@{\hspace{0.05cm}}|@{}c@{\hspace{0.05cm}}|@{}c@{\hspace{0.05cm}}|@{}c@{\hspace{0.05cm}}|@{}c@{\hspace{0.05cm}}|@{}c@{\hspace{0.05cm}}|@{}c@{\hspace{0.05cm}}|@{}c@{\hspace{0.05cm}}|@{}c@{\hspace{0.05cm}}}
            \Xhline{1pt}
			\multirow{2}{*}{Methods} & \multicolumn{10}{c}{Metric} \\
            \Xcline{2-11}{0.1pt}
			& Recall@20 & Recall@40 & Recall@60 & Recall@80 & Recall@100 & NDCG@20 & NDCG@40 & NDCG@60 & NDCG@80 & NDCG@100\\
            \Xhline{0.5pt}
	        NPA     &18.21 &28.84 &37.64 &50.86 &60.46 &7.04 &9.13 &10.68 &12.83 &14.39 \\	
			NAML    &18.49 &29.26 &36.83 &49.27 &60.95 &6.55 &8.73 &10.07 &12.08 &13.85  \\	
			NRMS    &18.97 &26.88 &38.83 &47.27 &60.59 &7.48 &9.26 &11.33 &12.69 &14.47  \\	
			CNE-SUE &17.68 &28.44 &36.85 &48.96 &61.01 &7.12 &9.23 &10.78 &12.73 &14.32\\
                UNBERT  &12.89 &24.08 &35.21 &43.85 &56.03 &4.53 &6.78 &8.77  &10.34 &12.07\\
                MINER   &15.20 &27.65 &37.87 &48.40 &60.83 &5.63 &8.15 &10.12 &11.68 &13.58\\
            \Xhline{0.5pt}
			JobFormer &\bf19.96 &\bf32.71 &\bf41.96 &\bf52.67 &\bf64.16 &\bf7.53 &\bf10.05 &\bf11.79 &\bf13.83 &\bf15.04 \\		
            \Xhline{1pt}
	\end{tabular*}}
\end{table*}  

\subsection{JD Recall Performance}
Firstly, we verify the effectiveness of JobFormer in the JD recall task. We repeat the experiment on each method 5 times and show results in Table \ref{tab:tab1}. Referring to this table, we have several findings: 1) Traditional word embedding-based recommendation methods outperform BERT-based pre-training recommendation methods. This indicates that word embedding can better encode the JD, mainly due to the JD structure and the relatively small-scale of our dataset. Thereby, we introduce the word embedding and TextCNN to encode JD texts in model design. 2) NRMS performs the best among all the word embedding-based methods on most criteria. The reason lies in that NRMS adopts deep models such as Transformer to learn contextual representations of JD items by capturing their interactions. This phenomenon indicates that multi-head self-attention can more effectively mine JD semantic information. 3) JobFormer achieves the best performance compared with other baselines on all criteria, e.g., JobFormer exceeds NRMS 2.31\% on Recall@$20$, 1.18\% on NDCG@$20$. The results validate the effectiveness of designed modules (e.g., local-global transformer) for processing JD.

\subsection{JD Ranking Performance}
Next, we further carry out the JD ranking task to discover user-interesting JDs from candidate JDs. In detail, we rank the candidate JDs according to the predicted click scores between candidate JD global representation and user profiles (i.e., personal skill distribution) for a personalized job recommendation. The ranking performance of different comparison methods is shown in Table \ref{tab:tab2}. We acquire a similar conclusion to the JD recall task that JobFormer achieves the best performance compared with other baselines. Note that for CTR ranking methods (i.e., FRNet and MaskNet), we utilize the raw samples of candidate JDs recalled by JobFormer to predict the click scores. JobFormer continues to perform best, mainly because JD representations obtain rich intra-job and inter-job information via semantic-enhanced transformer in the recall stage.

\begin{figure*}[!htb]
	\centering
	\includegraphics[width=185mm]{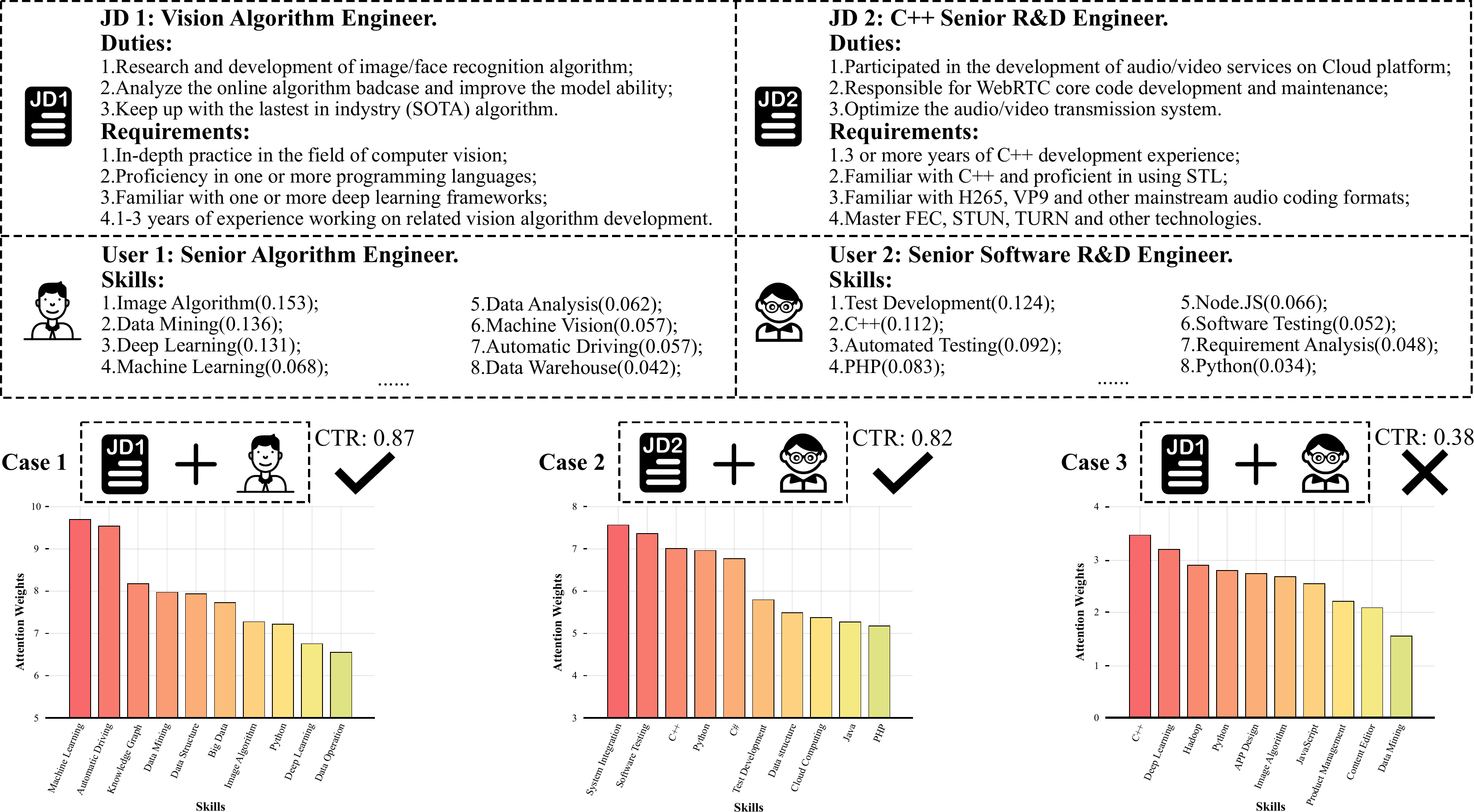}
	\caption{(Best viewed in color when zoomed in.) The example of interpretable CTR prediction.}\label{fig:f9}
\end{figure*}

\subsection{Ablation Study}
To verify the effectiveness of each module, we conduct more ablation studies, including: 1) LSTM+ and TextCNN+, we utilize the LSTM/TextCNN as item-level encoder, and then input the item representations to the transformer encoder, which is trained with KL divergence as the loss function. 2) local-global+, we adopt the TextCNN as item-level encoder, and then input the item representations to the semantic-enhanced transformer encoder with local-global attention, which is trained with KL divergence as the loss function. 3) w/o $M$, we remove the skill correlation enhancement $M(\cdot)$ in JobFormer. 4) w/o $R$, we remove the relation consistency $R(\cdot)$ in JobFormer. The bottom section of Table \ref{tab:tab1} records the results, which reveal that: 1) TextCNN+ performs better than LSTM+, which verifies that TextCNN can better model the JD items considering that TextCNN can process an unfixed-length sequence to capture more information~\cite{KalchbrennerGB14}, but the LSTM usually requires a fixed-length sentence. 2) local-global+ can improve performance, which indicates that local-global attention can effectively model the intra-job and inter-job information. 3) w/o $M$ performs superior to the local-global+, which shows that the relation consistency can further improve the recall performance. 4) w/o $R$ performs best in variants, which indicates that it is crucial to consider skill correlations. 5) JobFormer performs best compared with all the variants, which reveals that all the designed modules can promote the prediction of skill distribution in the recall stage.

\subsection{Parameter Analysis}
\textbf{Influence of Local-Global Heads.}  Considering that JobFormer with various local heads and global heads can have different emphasis on intra-job information and inter-job information, we fix the total heads as $N = 8$ and tune the local-global heads in $\{(N_l=4, N_g=4), (N_l=5, N_g=3), (N_l=6, N_g=2), (N_l=7, N_g=1)\}$, to empirically investigate the impact on JD recall performance. The top section of Table \ref{tab:tab3} depicts the results, the performance (including Recall and NDCG) of JobFormer firstly increases and then decreases. The JobFormer acquires best performance when local-global is $(N_l=6, N_g=2)$. This phenomenon confirms that the intra-job information is essential for learning skill-aware representation, but the neighbor JD can also provide additional supplementary information, i.e., the performance of $(N_l=7, N_g=1)$ is worse than $(N_l=6, N_g=2)$. Therefore, we need proper attention to neighbor information.

\noindent\textbf{Number of Neighbors.} To validate the influence of neighbors on the JD representation learning, we incorporate neighbors with different numbers (i.e., $L \in \{1, 2, 3, 4, 5\}$) to empirically investigate the impact of neighbors on JD recall task. Note that we fix the local-global heads as $(N_l=6, N_g=2)$. The bottom section of Table \ref{tab:tab3} depicts the results, the performance of JobFormer also increases firstly, and then decreases on various criteria. The reason is that more neighbors can even bring noise and the over-smooth problem.

\noindent\textbf{Influence of Balance Parameters.} To explore the influence of hyper-parameters, we tune the $\lambda, \mu \in \{0.1, 0.2, 0.4, 0.6, 0.8\}$ to conduct more experiments. The top section of Figure \ref{fig:f3} depicts the performance of different $\lambda$. With the increase of $\lambda$, the results of JD recall first increase and then decrease. This shows that skill correlation enhancement has a promoting effect, but over-considering skill correlation may introduce bias for prediction. The bottom section of Figure \ref{fig:f3} depicts the performance with different $\mu$. The results of JD recall first increase with the increase of $\mu$, and then decrease after $\mu > 0.4$. This shows that relation consistency actually has a promoting effect, but the label distribution prediction loss (i.e., KL-divergence) still has more contributions to model learning.

\subsection{Results on Public Dataset}
Furthermore, we also evaluate JobFormer on public dataset, i.e., SBU 3DFE dataset~\cite{yin20063d}. The SBU 3DFE database contains 2,500 facial
expression images. A 243-dimensional feature vector is extracted from each image by the method of Local Binary Patterns (LBP). Each image is scored by 23 persons on the 6 basic emotion labels (i.e., happiness, sadness, surprise, fear, anger, and disgust) with a 5-level scale. We score and normalize them into a label distribution over all the 6 emotion labels following~\cite{yin20063d}. Table \ref{tab:tab4} records the results, we can acquire the similar analyses as private dataset that JobFormer can also achieve the best recall performance. This phenomenon validates the generalization of JobFormer. 


\subsection{Case Study}

\textbf{JD Recall Visualization.} Firstly, we evaluate whether JobFormer could effectively recall user-related JDs from a large-scale JD pool. Figure \ref{fig:f7} shows the recall results of top-3 JDs given the corresponding user query. We find that most of the recalled candidate JDs according to the cosine similarity are exactly matched (title displayed in red font) with the ground-truth. Other outputs are also reasonable. Using the first row of Figure \ref{fig:f7} as an example, the 2nd and 3rd candidate JDs also have corresponding keywords such as “front-end”, “W3C standards” are related to skill “HTML5”, and “Spring MVC”, “MyBatis” are related to skill “Java”. Meanwhile, other skills (e.g., “MySQL”) are also required for Front-end Development Engineer and Java R\&D Engineer. Besides, we also exhibit the results of the state-of-the-art recommendation model NRMS, which is inferior to JobFormer that it only matches the correct JD on 2nd or 3rd location.

\noindent\textbf{Interpretable CTR Prediction.} To verify whether JobFormer could highlight the most critical skills that have strong contributions to CTR prediction, we present the attention weights of three different JDs with their corresponding user profiles, the results are recorded in Figure \ref{fig:f9}. In detail, we take the cumulative attention weights of each dimension of $[CLS]$ token as the level of emphasis the entire JD places on personal skills. Considering the page limitation, we only show the top-10 skills for interpretability analysis.

From the figures, we find that: 1) the skill-aware JD representation can accurately capture most of the relevant skills. For example, the skills “Machine Learning”, “Automatic Driving”, “Data Mining”, “Image Algorithm” and “Deep Learning” directly correspond to the skills of user1, which have the higher attention weights (i.e., 9.61, 9.54, 7.94, 7.19, 6.78). Other skills are also highly relevant to Vision Algorithm Engineer. Furthermore, the predicted click score (0.87) means that JobFormer is more likely to recommend JD1 to user1. 2) JobFormer can further efficiently identify JDs that mismatch with the current user. For example, the predicted click score in Case3 (0.38) is much lower than Case1 and Case2, because JD1 is intended for a computer vision position, while user2 is involved in the software development industry. Therefore, the obtained skill attention weights are relatively low.

\begin{figure}[!htb]
	\centering
	\includegraphics[width=85mm]{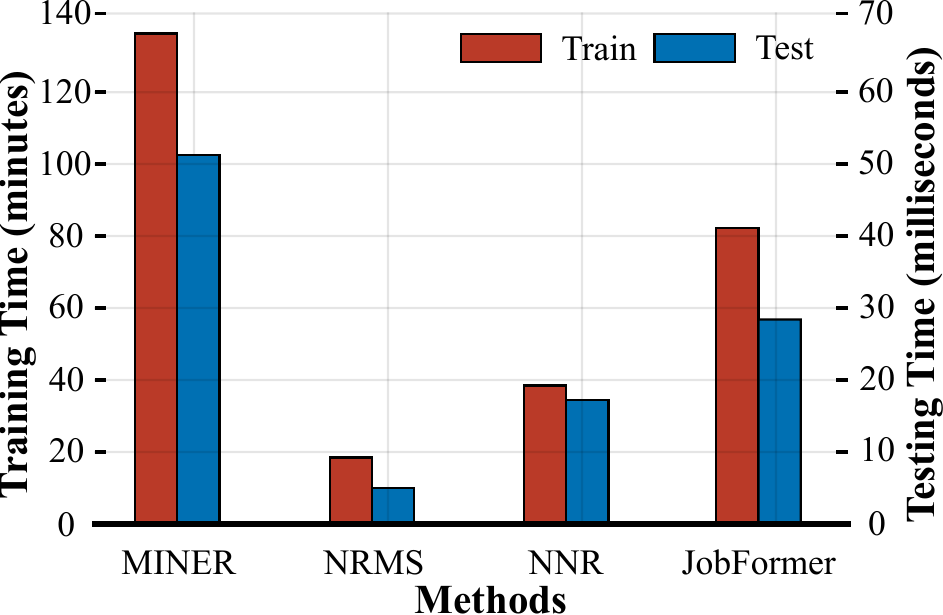}
	\caption{The training and testing efficiency of JobFormer and compared deep models on JD recall task.}\label{fig:f8}
\end{figure}

\subsection{Computational Efficiency} 
To evaluate the efficiency, we exhibit the computational times of JobFormer and compared deep models on JD recall task. Specifically, all of our experiments are conducted on a server with 2-core CPU@2.40GHz, 160GB RAM, and a NVIDIA Tesla V100 SXM2 GPU. As shown in Figure~\ref{fig:f8}, we observe that the training time is 82 minutes, which is much shorter than the comparison BERT-based recommendation method MINER. Although the training time of our method is slightly higher than NRMS and NNR, our performance outperforms these two methods. Moreover, after the training process, the average cost of each instance in testing set is 28ms. It clearly validates that our model can be effectively used in the real-world management analysis system.


\section{Conclusion}
In this paper, we propose a skill-aware representation method (JobFormer) that can utilize job descriptions and user profiles (personal skill distribution) to accomplish personalized job recommendation. Our method contains a recall stage and a ranking stage. In the recall stage, we first leverage semantic-enhanced transformer to parse JDs and guide the representation learning of JD via personal skill distribution. In detail, we design an encoder with the local-global attention mechanism to mine the intra-job and inter-job dependencies from JD tuples. With the skill-aware JD representation, we can recall a portion of JDs relevant to the user as a candidate set from a large-scale JD pool according to the JD-user cosine similarity. In the ranking stage, candidate JDs are further computed with user profiles for CTR prediction and ranked for personalized JD recommendation. Experiments on real-world and public datasets can well demonstrate the effectiveness and interpretability of JobFormer.



\ifCLASSOPTIONcompsoc
  \section*{Acknowledgment}
National Key RD Program of China (2022YFF0712100),  NSFC (61906092, 62006118, 62276131),  Natural Science Foundation of Jiangsu Province of China under Grant (BK20200460), Jiangsu Shuangchuang (Mass Innovation and Entrepreneurship) Talent Program. Young Elite Scientists Sponsorship Program by CAST, CAAI-Huawei MindSpore Open Fund (CAAIXSJLJJ-2021-014B), the Fundamental Research Funds for the Central Universities (NO.NJ2022028, No.30922010317)

\ifCLASSOPTIONcaptionsoff
  \newpage
\fi

\bibliographystyle{IEEEtranN}{\small
	\bibliography{acmart}}



\vspace{-1.2cm}
\begin{IEEEbiography}[{\includegraphics[width=1in,height=1.25in,clip,keepaspectratio]{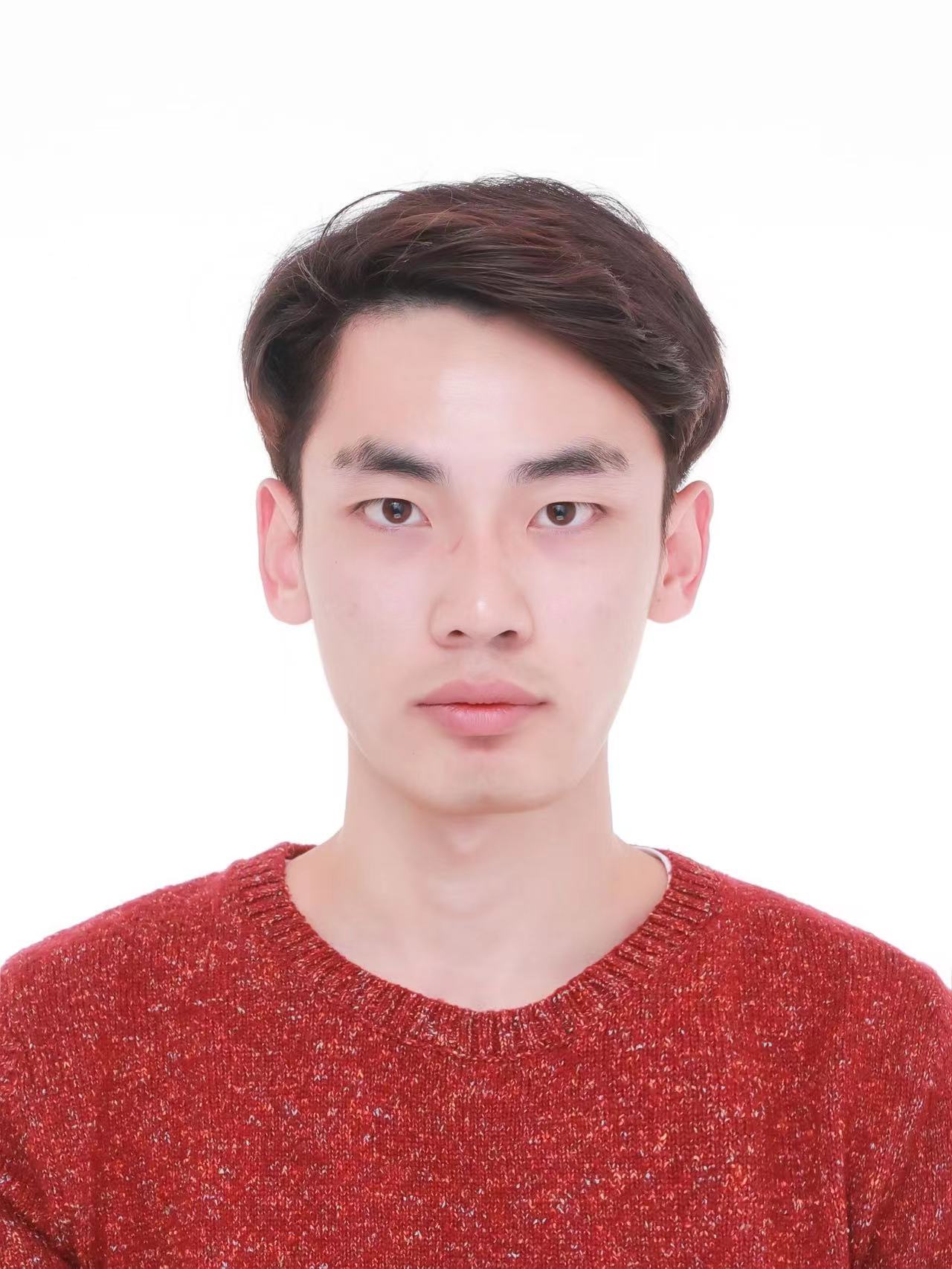}}]{Zhihao Guan}
is working towards the Ph.D. degree with the School of Computer Science and Engineering, in Nanjing University of Science and Technology, China. His research interests lie primarily in deep learning and data mining, including cross-modal learning.
\end{IEEEbiography}
\vspace{-1.2cm}
\begin{IEEEbiography}[{\includegraphics[width=1in,height=1.25in,clip,keepaspectratio]{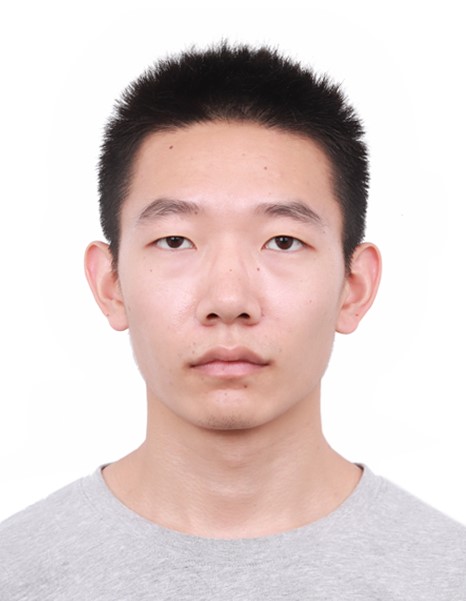}}]{Jia-Qi Yang}
	earned his M.E. degree in 2021 and is currently pursuing a Ph.D. at the State Key Laboratory for Novel Software Technology, Nanjing University, China. His research are primarily focused on machine learning and data mining, with specific expertise in uncertainty calibration, recommendation systems, and AI for science. He serves as a Program Committee member and Reviewer for prestigious conferences like AAAI, NeurIPS, ICLR, and others.
\end{IEEEbiography}
\vspace{-1.2cm}
\begin{IEEEbiography}[{\includegraphics[width=1in,height=1.25in,clip,keepaspectratio]{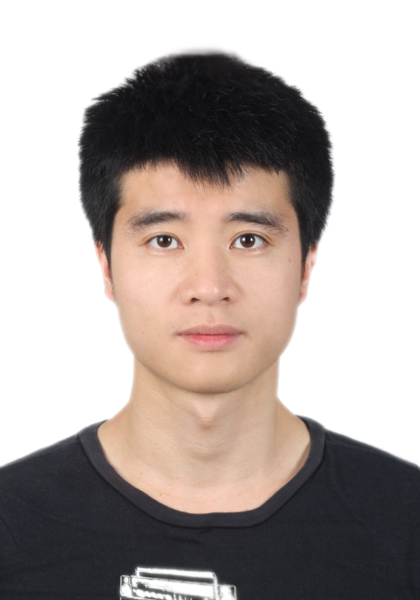}}]{Yang Yang}
received the Ph.D. degree in computer science, Nanjing University, China in 2019. At the same year, he became a faculty member at Nanjing University of Science and Technology, China. He is currently a Professor with the school of Computer Science and Engineering. His research interests lie primarily in machine learning and data mining, including heterogeneous learning, model reuse, and incremental mining. He serves as PC in leading conferences such as IJCAI, AAAI, ICML, NIPS, etc.
\end{IEEEbiography}
\vspace{-1.2cm}
\begin{IEEEbiography}[{\includegraphics[width=1in,height=1.25in,clip,keepaspectratio]{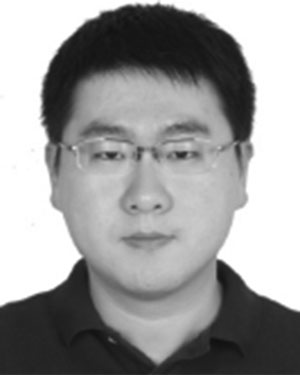}}]{Hengshu Zhu}(SM’19)
	is currently a principal data scientist $\&$ architect at Baidu Inc. He received the Ph.D. degree in 2014 and B.E. degree in 2009, both in Computer Science from University of Science and Technology of China (USTC), China. His general area of research is data mining and machine learning, with a focus on developing advanced data analysis techniques for innovative business applications. He has published prolifically in refereed journals and conference proceedings, including IEEE Transactions on Knowledge and Data Engineering (TKDE), IEEE Transactions on Mobile Computing (TMC), ACM Transactions on Information Systems (ACM TOIS), ACM Transactions on Knowledge Discovery from Data (TKDD), ACM SIGKDD, ACM SIGIR, WWW, IJCAI, and AAAI. He has served regularly on the organization and program committees of numerous conferences, including as a program co-chair of the KDD Cup-2019 Regular ML Track, and a founding co-chair of the first International Workshop on Organizational Behavior and Talent Analytics (OBTA) and the International Workshop on Talent and Management Computing (TMC), in conjunction with ACM SIGKDD. He was the recipient of the Distinguished Dissertation Award of CAS (2016), the Distinguished Dissertation Award of CAAI (2016), the Special Prize of President Scholarship for Postgraduate Students of CAS (2014), the Best Student Paper Award of KSEM-2011, WAIM-2013, CCDM-2014, and the Best Paper Nomination of ICDM-2014. He is the senior member of IEEE, ACM, and CCF. 
\end{IEEEbiography}
\vspace{-1.2cm}
\begin{IEEEbiography}[{\includegraphics[width=1in,height=1.25in,clip,keepaspectratio]{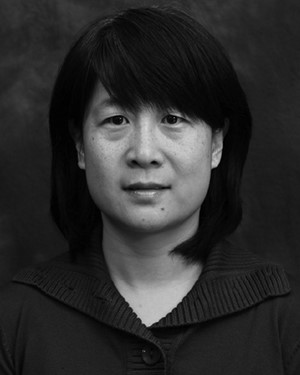}}]{Wenjie Li}
received the PhD degree in systems engineering and engineering management from the Chinese University of Hong Kong, Hong Kong, in 1997. She is currently an associate professor with the Department of Computing, The Hong Kong Polytechnic University. Her main research interests include online social network analysis, natural language processing, and document summarization.
\end{IEEEbiography}
\vspace{-1.2cm}
\begin{IEEEbiography}[{\includegraphics[width=1in,height=1.25in,clip,keepaspectratio]{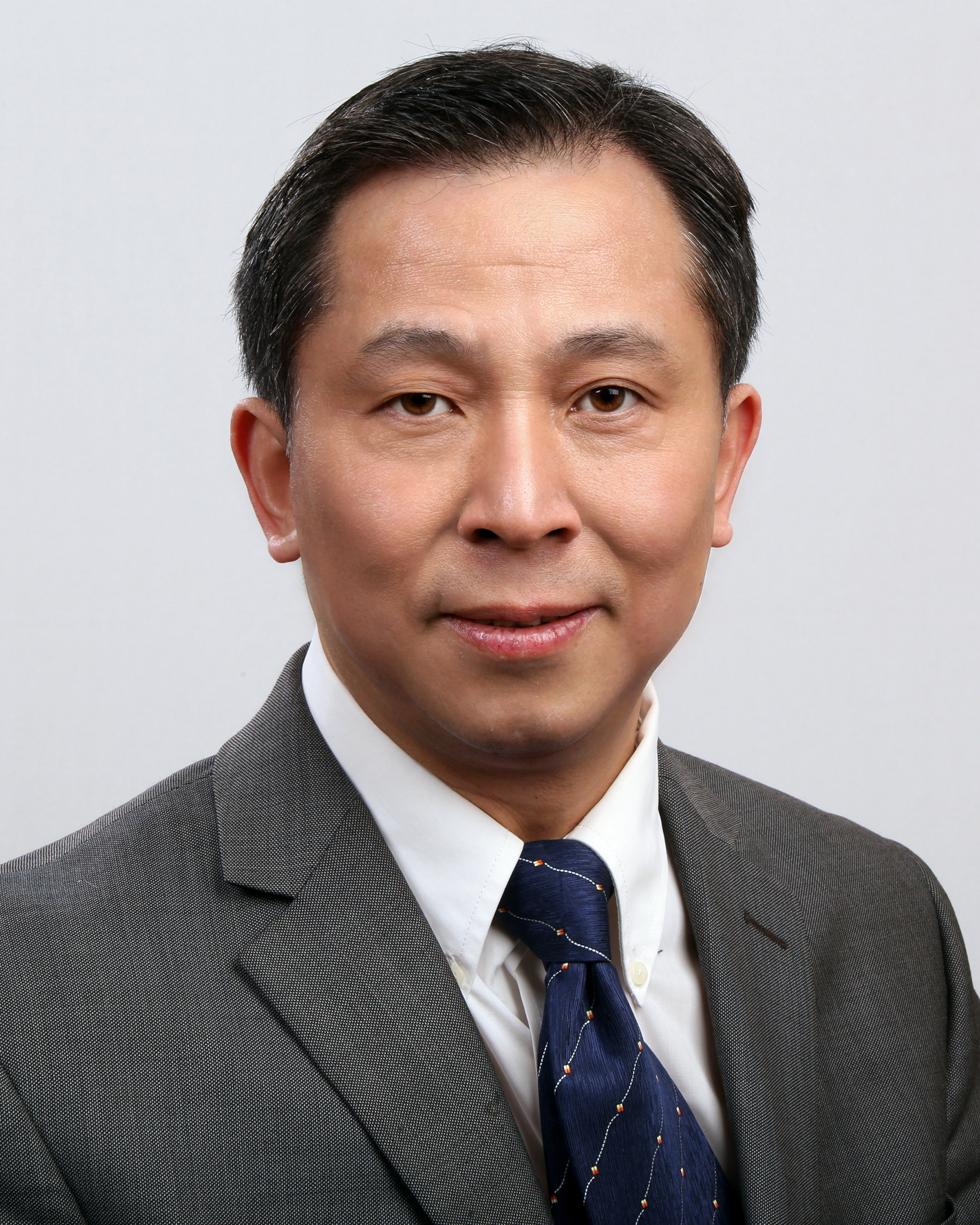}}]{Hui Xiong} (Fellow'20)
is a Professor at the Artificial Intelligence Thrust, The Hong Kong University of Science and Technology. Xiong’s research interests include data mining, mobile computing, and their applications in business. Xiong received his PhD in Computer Science from University of Minnesota, USA. He has served regularly on the organization and program committees of numerous conferences, including as a Program Co-Chair of the Industrial and Government Track for the 18th ACM SIGKDD International Conference on Knowledge Discovery and Data Mining (KDD), a Program Co-Chair for the IEEE 2013 International Conference on Data Mining (ICDM), a General Co-Chair for the 2015 IEEE International Conference on Data Mining (ICDM), and a Program Co-Chair of the Research Track for the 2018 ACM SIGKDD International Conference on Knowledge Discovery and Data Mining. He received the 2021 AAAI Best Paper Award and the 2011 IEEE ICDM Best Research Paper award. For his outstanding contributions to data mining and mobile computing, he was elected an AAAS Fellow and an IEEE Fellow in 2020.
\end{IEEEbiography}
\vspace{-1.2cm}
\end{document}



\title{The supplementary of Learning Adaptive Embedding Considering Incremental Class}

\author{Yang Yang,~\IEEEmembership{}
	    Zhen-Qiang Sun,~\IEEEmembership{}
	    Hengshu Zhu,~\IEEEmembership{Senior Member, IEEE}
	    Yanjie Fu,~\IEEEmembership{}
        Hui Xiong,~\IEEEmembership{Senior Member, IEEE}
        and Jian Yang,~\IEEEmembership{Member, IEEE}
}


\maketitle
\begin{table*}[!htb]{
		\centering
		\caption{Classification of known classes and novel class on three real-world datasets (single novel class cases). The best results are highlighted in bold.}
		\label{tab:tab1}
		\begin{tabular*}{1\textwidth}{@{\extracolsep{\fill}}@{}l|c|c|c|c|c|c|c|c}
			\toprule
			\multirow{2}{*}{Methods} & \multicolumn{4}{c|}{ \tabincell{c}{Average NA $\uparrow$}} & \multicolumn{4}{c}{\tabincell{c}{ Average Macro-F-Measure $\uparrow$ }}\\
			\cmidrule(l){2-9}
			& OTTO              & SNSR           & Tiny-ImageNet &ImageNet      & OTTO              & SNSR              & Tiny-ImageNet &ImageNet \\
			\midrule
			Iforest  &     .053$\pm$.002 &     .074$\pm$.006 &     .038$\pm$.004 & .031$\pm$.001 &     .072$\pm$.005 &     .081$\pm$.012 &     .033$\pm$.002 &.029$\pm$.002 \\
			One-SVM  &     .049$\pm$.002 &     .089$\pm$.002 &     .035$\pm$.002 & .030$\pm$.001&     .078$\pm$.002 &     .080$\pm$.007 &     .033$\pm$.004 &.029$\pm$.005\\
			LACU-SVM &     .043$\pm$.002 &     .061$\pm$.005 &     .029$\pm$.004 & .025$\pm$.004&     .046$\pm$.004 &     .068$\pm$.006 &     .025$\pm$.003 &.022$\pm$.004\\
			SENC-MAS &     .035$\pm$.003 &     .066$\pm$.011 &     .026$\pm$.005 & .024$\pm$.003&     .045$\pm$.004 &     .044$\pm$.004 &     .030$\pm$.002 &.025$\pm$.002\\
			\midrule
			ODIN-CNN &     .093$\pm$.004 &     .051$\pm$.012 &     .043$\pm$.002 &.037$\pm$.002 &     .091$\pm$.006 &     .091$\pm$.005 &     .051$\pm$.003 &.040$\pm$.002\\
			CFO      &     .086$\pm$.004 &     .114$\pm$.010 &     .030$\pm$.003 &.025$\pm$.004 &     .087$\pm$.001 &     .129$\pm$.006 &     .039$\pm$.003 &.033$\pm$.001\\
			CPE      &     .135$\pm$.011 &     .185$\pm$.008 &     .053$\pm$.003 &.044$\pm$.003 &     .111$\pm$.002 &     .102$\pm$.006 &     .073$\pm$.005 &.061$\pm$.002\\
			\midrule
			DEC      &     .077$\pm$.011 &     .112$\pm$.011 &     .023$\pm$.003 &.016$\pm$.002 &     .051$\pm$.005 &     .087$\pm$.008 &     .033$\pm$.003 &.025$\pm$.002\\
			\midrule
			CILF     & \bf .140$\pm$.003 & \bf .198$\pm$.012 & \bf .054$\pm$.003 &\bf .049$\pm$.002 & \bf .119$\pm$.010 & \bf .180$\pm$.008 & \bf .076$\pm$.003 &\bf .067$\pm$.001\\
		\end{tabular*}
		\begin{tabular*}{1\textwidth}{@{\extracolsep{\fill}}@{}l|c|c|c|c|c|c|c|c}
			\toprule
			\multirow{2}{*}{Methods} & \multicolumn{4}{c|}{\tabincell{c}{Average Micro-F-Measure $\uparrow$ }} & \multicolumn{4}{c}{\tabincell{c}{Average AUROC $ \uparrow$ }}\\
			\cmidrule(l){2-9}
			& OTTO              & SNSR              & Tiny-ImageNet   &ImageNet   & OTTO              & SNSR              & Tiny-ImageNet  &ImageNet\\
			\midrule
			Iforest  &     .064$\pm$.004 &     .089$\pm$.002 &     .043$\pm$.003 &.035$\pm$.003 &     .078$\pm$.009 &     .127$\pm$.002 &     .092$\pm$.005 &.066$\pm$.002\\
			One-SVM  &     .065$\pm$.002 &     .087$\pm$.004 &     .042$\pm$.001 &.037$\pm$.002 &     .078$\pm$.005 &     .138$\pm$.003 &     .078$\pm$.005 &.061$\pm$.002\\
			LACU-SVM &     .063$\pm$.004 &     .062$\pm$.003 &     .039$\pm$.002 &.034$\pm$.002 &     .074$\pm$.010 &     .095$\pm$.005 &     .072$\pm$.003 &.053$\pm$.003\\
			SENC-MAS &     .057$\pm$.004 &     .058$\pm$.003 &     .037$\pm$.004 &.035$\pm$.003 &     .064$\pm$.011 &     .087$\pm$.010 &     .072$\pm$.003 &.054$\pm$.001\\
			\midrule
			ODIN-CNN &     .054$\pm$.005 &     .067$\pm$.011 &     .047$\pm$.002 &.036$\pm$.001 &     .118$\pm$.006 &     .129$\pm$.002 &     .118$\pm$.006 &.092$\pm$.003\\
			CFO      &     .071$\pm$.007 &     .182$\pm$.006 &     .034$\pm$.002 &.034$\pm$.001 &     .109$\pm$.005 &     .137$\pm$.005 &     .094$\pm$.002 &.078$\pm$.003\\
			CPE      & \bf .089$\pm$.003 &     .146$\pm$.003 &     .089$\pm$.003 &.068$\pm$.002 &     .142$\pm$.004 &     .177$\pm$.005 &     .126$\pm$.003 &.094$\pm$.001\\
			\midrule
			DEC      &     .031$\pm$.002 &     .070$\pm$.007 &     .031$\pm$.002 &.027$\pm$.007 &     .104$\pm$.004 &     .102$\pm$.006 &     .096$\pm$.003 &.076$\pm$.001\\
			\midrule
			CILF     &     .085$\pm$.005 & \bf .211$\pm$.013 & \bf .101$\pm$.005 &\bf 071$\pm$.002 & \bf .160$\pm$.014 & \bf .204$\pm$.012 & \bf .135$\pm$.005 &\bf .103$\pm$.002\\
			\bottomrule
	\end{tabular*}}
\end{table*}

\begin{table*}[!htb]{
		\centering
		\caption{Classification of known classes and novel class on three real-world datasets (multiple novel class cases). The best results are highlighted in bold.}
		\label{tab:tab3}
		\begin{tabular*}{1\textwidth}{@{\extracolsep{\fill}}@{}l|c|c|c|c|c|c|c|c}
			\toprule
			\multirow{2}{*}{Methods} & \multicolumn{4}{c|}{\tabincell{c}{Average NA $\uparrow$ }} & \multicolumn{4}{c}{\tabincell{c}{Average Macro-F-Measure $\uparrow$ }}\\
			\cmidrule(l){2-9}
			& OTTO              & SNSR              & Tiny-ImageNet   &ImageNet  & OTTO              & SNSR              & Tiny-ImageNet &ImageNet\\
			\midrule
			Iforest  &     .081$\pm$.019 &     .110$\pm$.025 &     .077$\pm$.035 &.064$\pm$.025 &     .114$\pm$.016 &     .158$\pm$.003 &     .072$\pm$.037 &.058$\pm$.031\\
			One-SVM  &     .082$\pm$.033 &     .123$\pm$.012 &     .092$\pm$.038 &.079$\pm$.038 &     .136$\pm$.030 &     .129$\pm$.035 &     .079$\pm$.029 &.061$\pm$.036\\
			LACU-SVM &     .087$\pm$.035 &     .122$\pm$.022 &     .042$\pm$.018 &.038$\pm$.022 &     .098$\pm$.017 &     .120$\pm$.010 &     .056$\pm$.020 &.041$\pm$.025\\
			SENC-MAS &     .072$\pm$.023 &     .099$\pm$.037 &     .088$\pm$.028 &.086$\pm$.044 &     .094$\pm$.017 &     .068$\pm$.027 &     .080$\pm$.038 &.069$\pm$.033\\
			\midrule
			ODIN-CNN &     .122$\pm$.029 &     .120$\pm$.004 &     .104$\pm$.032 &.072$\pm$.025 &     .150$\pm$.034 &     .131$\pm$.038 &     .113$\pm$.031 &.084$\pm$.041\\
			CFO      &     .131$\pm$.049 &     .170$\pm$.037 &     .080$\pm$.044 &.065$\pm$.034 &     .137$\pm$.032 &     .202$\pm$.036 &     .088$\pm$.047 &.077$\pm$.023\\
			CPE      &     .161$\pm$.017 &     .212$\pm$.010 &     .105$\pm$.014 &.073$\pm$.020 &     .165$\pm$.043 &     .149$\pm$.035 &     .090$\pm$.017 &.079$\pm$.019\\
			\midrule
			DEC      &     .132$\pm$.043 &     .159$\pm$.003 &     .063$\pm$.025 &.077$\pm$.046 &     .109$\pm$.040 &     .128$\pm$.044 &     .111$\pm$.022 &.082$\pm$.010\\
			\midrule
			CILF     & \bf .184$\pm$.031 & \bf .244$\pm$.046 & \bf .124$\pm$.005 &\bf .089$\pm$.015 & \bf .166$\pm$.030 & \bf .226$\pm$.010 & \bf .125$\pm$.048 &\bf .086$\pm$.016\\
		\end{tabular*}
		\begin{tabular*}{1\textwidth}{@{\extracolsep{\fill}}@{}l|c|c|c|c|c|c|c|c}
			\toprule
			\multirow{2}{*}{Methods} & \multicolumn{4}{c|}{\tabincell{c}{Average Micro-F-Measure $\uparrow$ }} & \multicolumn{4}{c}{\tabincell{c}{Average AUROC $ \uparrow$}}\\
			\cmidrule(l){2-9}
			& OTTO              & SNSR              & Tiny-ImageNet  &ImageNet   & OTTO              & SNSR              & Tiny-ImageNet &ImageNet\\
			\midrule
			Iforest  &     .108$\pm$.051 &     .148$\pm$.005 &     .058$\pm$.027 &.047$\pm$.041 &     .131$\pm$.052 &     .170$\pm$.018 &     .141$\pm$.026 &.084$\pm$.056\\
			One-SVM  &     .098$\pm$.004 &     .172$\pm$.016 &     .085$\pm$.006 &.061$\pm$.036 &     .123$\pm$.024 &     .192$\pm$.041 &     .130$\pm$.033 &.092$\pm$.026\\
			LACU-SVM &     .107$\pm$.029 &     .092$\pm$.013 &     .095$\pm$.026 &.041$\pm$.025 &     .125$\pm$.037 &     .171$\pm$.034 &     .142$\pm$.028 &.075$\pm$.014\\
			SENC-MAS &     .117$\pm$.013 &     .098$\pm$.007 &     .089$\pm$.043 &.069$\pm$.033 &     .126$\pm$.015 &     .161$\pm$.028 &     .104$\pm$.016 &.072$\pm$.040\\
			\midrule
			ODIN-CNN &     .144$\pm$.005 &     .121$\pm$.015 &     .111$\pm$.033 &.085$\pm$.017 &     .144$\pm$.018 &     .201$\pm$.014 &     .152$\pm$.008 &.097$\pm$..015\\
			CFO      &     .135$\pm$.016 &     .239$\pm$.038 &     .093$\pm$.024 &.080$\pm$.022 &     .156$\pm$.049 &     .188$\pm$.022 &     .146$\pm$.025 &.095$\pm$.042\\
			CPE      &     .150$\pm$.031 &     .194$\pm$.017 &     .136$\pm$.048 &.082$\pm$.034 &     .212$\pm$.011 &     .226$\pm$.040 &     .164$\pm$.016 &.095$\pm$.028\\
			\midrule
			DEC      &     .104$\pm$.010 &     .115$\pm$.042 &     .082$\pm$.034 &.093$\pm$.028 &     .134$\pm$.024 &     .141$\pm$.015 &     .138$\pm$.042 &.104$\pm$.036\\
			\midrule
			CILF     & \bf .168$\pm$.026 & \bf .273$\pm$.040 & \bf .166$\pm$.028 &\bf .097$\pm$.021 & \bf .214$\pm$.055 & \bf .292$\pm$.012 & \bf .168$\pm$.031 &\bf .110$\pm$.025\\
			\bottomrule
	\end{tabular*}}
\end{table*}

In the supplementary materials, we mainly focus on giving more experimental analysis.

\section{Compared Methods}
To validate the effectiveness of proposed CILF, we compared it with existing state-of-the-art novel class detection approaches and incremental learning methods. 

First, we compared CILF with existing NCD and incremental NCD methods. Including traditional anomaly detection and linear methods: Iforest~\cite{LiuTZ08}, One-Class SVM (One-SVM)~\cite{ScholkopfPSSW01}, LACU-SVM (LACU)~\cite{DaYZ14}, SENC-MAS (SENC)~\cite{MuZDLZ17}; as well as deep methods: ODIN-CNN (ODIN)~\cite{LiangLS18}, CFO~\cite{NealOFWL18}, CPE~\cite{WangKCTK19} and DTC~\cite{HanVZ19}. Abbreviations in parentheses. Specifically,
\begin{itemize}
	\item Iforest: an ensemble tree method to detect outliers;
	\item One-Class SVM (One-SVM): a baseline for out-of-class detection and classification;
	\item LACU-SVM (LACU): a SVM-based method that incorporates the unlabeled data from open set for unknown class detection;
	\item SENC-MAS (SENC): a matrix sketching method that approximates original information with a dynamic low-dimensional structure;
	\item ODIN-CNN (ODIN): a CNN-based method that distinguishes in-distribution and out-of-distribution over softmax score;
	\item CFO: a generative method that adopts an encoder-decoder GAN to generate synthetic unknown instances;
	\item CPE: a CNN-based ensemble method, which adaptively updates the prototype for detection;
	\item DTC: an extended deep transfer clustering method for novel class detection.
\end{itemize}

To validate the incremental model update, we also compare CILF with state-of-the-art forgetting methods: DNN-Base, DNN-L2, DNN-EWC~\citep{KirkpatrickPRVD16}, IMM~\citep{LeeKJHZ17}, DEN~\citep{Jeongtae2017}, and each time window is regarded as a task for these methods. In detail, the compared methods are: 
\begin{itemize}
	\item {\bf DNN-Base:} Base DNN with $L_2$-regularizations;
	\item {\bf DNN-L2:} Base DNN, where at each stage t, $\Theta_t$ is initialized as $\Theta_{t-1}$ and continuously trained with $L_2$-regularization between $\Theta_t$ and $\Theta_{t-1}$;
	\item {\bf DNN-EWC:} Deep network trained with elastic weight consolidation for regularization, which remembers old stages by selectively slowing down learning on the weights important for those stages;
	\item {\bf IMM:} An incremental moment matching method with two extensions: Mean-IMM and Mode-IMM, which incrementally matches the posterior distribution of the neural network trained on the previous stages;
	\item {\bf DEN:} A deep network architecture for incremental learning, which can dynamically decide its network structure with a sequence of stages and learn the overlapping knowledge sharing structure among stages.
\end{itemize}

\begin{table}[htb]{
		\centering
		\caption{Forgetting measure of known classes on three real-world datasets (single novel class case). The best results are highlighted in bold.}
		\label{tab:tab2}
		\begin{tabular}{@{}l@{}|@{}c|@{}c|@{}c|@{}c}
			\hline
			\multirow{2}{*}{Methods} & \multicolumn{4}{c}{Forgetting $\downarrow$} \\ \cline{2-5} 
			& OTTO              & SNSR              & Tiny-ImageNet & ImageNet       \\
			\midrule
			Iforest  &     .056$\pm$.003 &     .045$\pm$.003 &     .064$\pm$.001 &.082$\pm$.003\\
			One-SVM  &     .054$\pm$.002 &     .049$\pm$.002 &     .066$\pm$.002 &.080$\pm$.001\\
			LACU-SVM &     .098$\pm$.002 &     .065$\pm$.001 &     .070$\pm$.001 &.085$\pm$.001\\
			SENC-MAS &     .078$\pm$.001 &     .061$\pm$.002 &     .072$\pm$.003 &.052$\pm$.002\\
			\midrule
			ODIN-CNN &     .034$\pm$.003 &     .029$\pm$.001 &     .023$\pm$.002 &.057$\pm$.003\\
			CFO      &     .045$\pm$.002 &     .028$\pm$.003 &     .029$\pm$.003 &.038$\pm$.002\\
			CPE      &     .040$\pm$.003 &     .017$\pm$.003 &     .019$\pm$.001 &.031$\pm$.001\\
			\midrule
			DEC      &     .055$\pm$.003 &     .027$\pm$.002 &     .032$\pm$.002 &.054$\pm$.002\\
			\midrule
			DNN-Base &     .047$\pm$.001 &     .029$\pm$.003 &     .051$\pm$.001 &.076$\pm$.002\\
			DNN-L2   &     .051$\pm$.002 &     .025$\pm$.002 &     .046$\pm$.003 &.071$\pm$.003\\
			DNN-EWC  &     .056$\pm$.003 &     .034$\pm$.002 &     .041$\pm$.002 &.071$\pm$.002\\
			IMM      &     .039$\pm$.001 &     .031$\pm$.001 &     .036$\pm$.001 &.065$\pm$.001\\
			DEN      &     .039$\pm$.002 &     .027$\pm$.003 &     .031$\pm$.001 &.063$\pm$.002\\
			\midrule
			CILF     & \bf .028$\pm$.001 & \bf .009$\pm$.001 & \bf .015$\pm$.002 &\bf .023$\pm$.002\\
			\bottomrule
	\end{tabular}}
\end{table}

\begin{table}[htb]{
		\centering
		\caption{Forgetting measure of known classes on three real-world datasets (multiple novel class case). The best results are highlighted in bold.}
		\label{tab:tab4}
		\begin{tabular}{@{}l@{}|@{}c|@{}c|@{}c|@{}c}
			\hline
			\multirow{2}{*}{Methods} & \multicolumn{4}{c}{Forgetting $\downarrow$} \\ \cline{2-5} 
			& OTTO              & SNSR              & Tiny-ImageNet     & ImageNet   \\
			\midrule
			Iforest  &     .049$\pm$.002 &     .041$\pm$.003 &     .051$\pm$.001 &.054$\pm$.004\\
			One-SVM  &     .058$\pm$.002 &     .054$\pm$.003 &     .057$\pm$.003 &.061$\pm$.003\\
			LACU-SVM &     .079$\pm$.001 &     .078$\pm$.002 &     .059$\pm$.002 &.056$\pm$.003\\
			SENC-MAS &     .070$\pm$.002 &     .054$\pm$.001 &     .063$\pm$.003 &.062$\pm$.002\\
			\midrule
			ODIN-CNN &     .047$\pm$.002 &     .038$\pm$.003 &     .034$\pm$.001 &.025$\pm$.005\\
			CFO      &     .059$\pm$.003 &     .024$\pm$.002 &     .023$\pm$.002 &.039$\pm$.006\\
			CPE      &     .031$\pm$.001 &     .025$\pm$.003 &     .023$\pm$.003 &.025$\pm$.002\\
			\midrule
			DEC      &     .045$\pm$.003 &     .043$\pm$.001 &     .039$\pm$.002 &.028$\pm$.003\\
			\midrule
			DNN-Base &     .044$\pm$.002 &     .043$\pm$.001 &     .048$\pm$.001 &.055$\pm$.010\\
			DNN-L2   &     .041$\pm$.003 &     .039$\pm$.003 &     .045$\pm$.002 &.062$\pm$.003\\
			DNN-EWC  &     .038$\pm$.001 &     .036$\pm$.002 &     .041$\pm$.002 &.050$\pm$.006\\
			IMM      &     .031$\pm$.003 &     .039$\pm$.001 &     .035$\pm$.003 &.057$\pm$.004\\
			DEN      &     .035$\pm$.001 &     .034$\pm$.002 &     .029$\pm$.002 &.042$\pm$.002\\
			\midrule
			CILF     & \bf .021$\pm$.002 & \bf .013$\pm$.003 & \bf .019$\pm$.001 &\bf .015$\pm$.003\\
			\bottomrule
	\end{tabular}}
\end{table} 

\begin{figure*}[!htb]
	\begin{center}
		\begin{minipage}[h]{28mm}
			\centering
			\includegraphics[width=28mm]{./fig/fig_tsne_single/1.pdf}\\
			\mbox{ ({\it a-1}) {Original-1}}
		\end{minipage}
		\begin{minipage}[h]{28mm}
			\centering
			\includegraphics[width=28mm]{./fig/fig_tsne_single/2.pdf}\\
			\mbox{ ({\it a-2}) {Original-2}}
		\end{minipage}
		\begin{minipage}[h]{28mm}
			\centering
			\includegraphics[width=28mm]{./fig/fig_tsne_single/3.pdf}\\
			\mbox{ ({\it a-3}) {Original-3}}
		\end{minipage}
		\begin{minipage}[h]{28mm}
			\centering
			\includegraphics[width=28mm]{./fig/fig_tsne_single/4.pdf}\\
			\mbox{ ({\it a-4}) {Original-4}}
		\end{minipage}
		\begin{minipage}[h]{28mm}
			\centering
			\includegraphics[width=28mm]{./fig/fig_tsne_single/5.pdf}\\
			\mbox{ ({\it a-5}) {Original-5}}
		\end{minipage}
		\begin{minipage}[h]{28mm}
			\centering
			\includegraphics[width=28mm]{./fig/fig_tsne_single/6.pdf}\\
			\mbox{ ({\it a-6}) {Original-6}}
		\end{minipage}
		
		\begin{minipage}[h]{28mm}
			\centering
			\includegraphics[width=28mm]{./fig/fig_tsne_single/7.pdf}\\
			\mbox{ ({\it b-1}) {CPE-1}}
		\end{minipage}
		\begin{minipage}[h]{28mm}
			\centering
			\includegraphics[width=28mm]{./fig/fig_tsne_single/8.pdf}\\
			\mbox{ ({\it b-2}) {CPE-2}}
		\end{minipage}
		\begin{minipage}[h]{28mm}
			\centering
			\includegraphics[width=28mm]{./fig/fig_tsne_single/9.pdf}\\
			\mbox{ ({\it b-3}) {CPE-3}}
		\end{minipage}
		\begin{minipage}[h]{28mm}
			\centering
			\includegraphics[width=28mm]{./fig/fig_tsne_single/10.pdf}\\
			\mbox{ ({\it b-4}) {CPE-4}}
		\end{minipage}
		\begin{minipage}[h]{28mm}
			\centering
			\includegraphics[width=28mm]{./fig/fig_tsne_single/11.pdf}\\
			\mbox{ ({\it b-5}) {CPE-5}}
		\end{minipage}
		\begin{minipage}[h]{28mm}
			\centering
			\includegraphics[width=28mm]{./fig/fig_tsne_single/12.pdf}\\
			\mbox{ ({\it b-6}) {CPE-6}}
		\end{minipage}
		
		\begin{minipage}[h]{28mm}
			\centering
			\includegraphics[width=28mm]{./fig/fig_tsne_single/13.pdf}\\
			\mbox{ ({\it c-1}) {DEC-1}}
		\end{minipage}
		\begin{minipage}[h]{28mm}
			\centering
			\includegraphics[width=28mm]{./fig/fig_tsne_single/14.pdf}\\
			\mbox{ ({\it c-2}) {DEC-2}}
		\end{minipage}
		\begin{minipage}[h]{28mm}
			\centering
			\includegraphics[width=28mm]{./fig/fig_tsne_single/15.pdf}\\
			\mbox{ ({\it c-3}) {DEC-3}}
		\end{minipage}
		\begin{minipage}[h]{28mm}
			\centering
			\includegraphics[width=28mm]{./fig/fig_tsne_single/16.pdf}\\
			\mbox{ ({\it c-4}) {DEC-4}}
		\end{minipage}
		\begin{minipage}[h]{28mm}
			\centering
			\includegraphics[width=28mm]{./fig/fig_tsne_single/17.pdf}\\
			\mbox{ ({\it c-5}) {DEC-5}}
		\end{minipage}
		\begin{minipage}[h]{28mm}
			\centering
			\includegraphics[width=28mm]{./fig/fig_tsne_single/18.pdf}\\
			\mbox{ ({\it c-6}) {DEC-6}}
		\end{minipage}
		\begin{minipage}[h]{28mm}
			\centering
			\includegraphics[width=28mm]{./fig/fig_tsne_single/19.pdf}\\
			\mbox{ ({\it d-1}) {CILF-1}}
		\end{minipage}
		\begin{minipage}[h]{28mm}
			\centering
			\includegraphics[width=28mm]{./fig/fig_tsne_single/20.pdf}\\
			\mbox{ ({\it d-2}) {CILF-2}}
		\end{minipage}
		\begin{minipage}[h]{28mm}
			\centering
			\includegraphics[width=28mm]{./fig/fig_tsne_single/21.pdf}\\
			\mbox{ ({\it d-3}) {CILF-3}}
		\end{minipage}
		\begin{minipage}[h]{28mm}
			\centering
			\includegraphics[width=28mm]{./fig/fig_tsne_single/22.pdf}\\
			\mbox{ ({\it d-4}) {CILF-4}}
		\end{minipage}
		\begin{minipage}[h]{28mm}
			\centering
			\includegraphics[width=28mm]{./fig/fig_tsne_single/23.pdf}\\
			\mbox{ ({\it d-5}) {CILF-5}}
		\end{minipage}
		\begin{minipage}[h]{28mm}
			\centering
			\includegraphics[width=28mm]{./fig/fig_tsne_single/24.pdf}\\
			\mbox{ ({\it d-6}) {CILF-6}}
		\end{minipage}
	\end{center}
	\caption{(Best view in color.) T-SNE Visualization for both known and unknown classes on CIFAR-10 in single novel class case. (a) original feature space; (b) Learned representations through single novel class detection method CPE~\cite{WangKCTK19}; (c) Learned representations through multiple novel class detection method DEC~\cite{HanVZ19}; (d) Learned representations through proposed CILF. Method$-t$ indicates the T-SNE of $t-$th time window of different methods.}\label{fig:f3}
\end{figure*}

\begin{figure*}[!htb]
	\begin{center}
		\begin{minipage}[h]{32mm}
			\centering
			\includegraphics[width=32mm]{./fig/fig_tsne_multiple/1.pdf}\\
			\mbox{ ({\it a-1}) {Original-1}}
		\end{minipage}
		\begin{minipage}[h]{32mm}
			\centering
			\includegraphics[width=32mm]{./fig/fig_tsne_multiple/2.pdf}\\
			\mbox{ ({\it a-2}) {Original-2}}
		\end{minipage}
		\begin{minipage}[h]{32mm}
			\centering
			\includegraphics[width=32mm]{./fig/fig_tsne_multiple/3.pdf}\\
			\mbox{ ({\it a-3}) {Original P3}}
		\end{minipage}
		\begin{minipage}[h]{32mm}
			\centering
			\includegraphics[width=32mm]{./fig/fig_tsne_multiple/4.pdf}\\
			\mbox{ ({\it a-4}) {Original-4}}
		\end{minipage}
		\begin{minipage}[h]{32mm}
			\centering
			\includegraphics[width=32mm]{./fig/fig_tsne_multiple/5.pdf}\\
			\mbox{ ({\it a-5}) {Original-5}}
		\end{minipage}		
		\begin{minipage}[h]{32mm}
			\centering
			\includegraphics[width=32mm]{./fig/fig_tsne_multiple/6.pdf}\\
			\mbox{ ({\it b-1}) {CPE-1}}
		\end{minipage}
		\begin{minipage}[h]{32mm}
			\centering
			\includegraphics[width=32mm]{./fig/fig_tsne_multiple/7.pdf}\\
			\mbox{ ({\it b-2}) {CPE-2}}
		\end{minipage}
		\begin{minipage}[h]{32mm}
			\centering
			\includegraphics[width=32mm]{./fig/fig_tsne_multiple/8.pdf}\\
			\mbox{ ({\it b-3}) {CPE-3}}
		\end{minipage}
		\begin{minipage}[h]{32mm}
			\centering
			\includegraphics[width=32mm]{./fig/fig_tsne_multiple/9.pdf}\\
			\mbox{ ({\it b-4}) {CPE-4}}
		\end{minipage}
		\begin{minipage}[h]{32mm}
			\centering
			\includegraphics[width=32mm]{./fig/fig_tsne_multiple/10.pdf}\\
			\mbox{ ({\it b-5}) {CPE-v5}}
		\end{minipage}
		
		\begin{minipage}[h]{32mm}
			\centering
			\includegraphics[width=32mm]{./fig/fig_tsne_multiple/11.pdf}\\
			\mbox{ ({\it c-1}) {DEC-1}}
		\end{minipage}
		\begin{minipage}[h]{32mm}
			\centering
			\includegraphics[width=32mm]{./fig/fig_tsne_multiple/12.pdf}\\
			\mbox{ ({\it c-2}) {DEC-2}}
		\end{minipage}
		\begin{minipage}[h]{32mm}
			\centering
			\includegraphics[width=32mm]{./fig/fig_tsne_multiple/13.pdf}\\
			\mbox{ ({\it c-3}) {DEC-v3}}
		\end{minipage}
		\begin{minipage}[h]{32mm}
			\centering
			\includegraphics[width=32mm]{./fig/fig_tsne_multiple/14.pdf}\\
			\mbox{ ({\it c-4}) {DEC-4}}
		\end{minipage}
		\begin{minipage}[h]{32mm}
			\centering
			\includegraphics[width=32mm]{./fig/fig_tsne_multiple/15.pdf}\\
			\mbox{ ({\it c-5}) {DEC-5}}
		\end{minipage}		
		\begin{minipage}[h]{32mm}
			\centering
			\includegraphics[width=32mm]{./fig/fig_tsne_multiple/16.pdf}\\
			\mbox{ ({\it d-1}) {CILF-1}}
		\end{minipage}
		\begin{minipage}[h]{32mm}
			\centering
			\includegraphics[width=32mm]{./fig/fig_tsne_multiple/17.pdf}\\
			\mbox{ ({\it d-2}) {CILF-2}}
		\end{minipage}
		\begin{minipage}[h]{32mm}
			\centering
			\includegraphics[width=32mm]{./fig/fig_tsne_multiple/18.pdf}\\
			\mbox{ ({\it d-3}) {CILF-3}}
		\end{minipage}
		\begin{minipage}[h]{32mm}
			\centering
			\includegraphics[width=32mm]{./fig/fig_tsne_multiple/19.pdf}\\
			\mbox{ ({\it d-4}) {CILF-4}}
		\end{minipage}
		\begin{minipage}[h]{32mm}
			\centering
			\includegraphics[width=32mm]{./fig/fig_tsne_multiple/20.pdf}\\
			\mbox{ ({\it d-5}) {CILF-5}}
		\end{minipage}	
	\end{center}
	\caption{T-SNE Visualization for both known and unknown classes on CIFAR-10 in multiple novel class case. (a) original feature space; (b) Learned representations through single detection method CPE~\cite{WangKCTK19}; (c) Learned representations through multi detection method DEC~\cite{HanVZ19}; (d) Learned representations through proposed CILF. Method$-t$ indicates the T-SNE of $t-$th time window of different methods.}\label{fig:f5}
\end{figure*}

\section{Complex Streaming Datasets}
We have conducted more experiments on datasets from other domains, i.e., OTTO and SNSR, and added two more challenging image datasets, i.e., Tiny-ImageNet and ImageNet. Specifically, the OTTO dataset from Kaggle and the SNSR dataset from UCI repository are chosen from problem sets of anomaly detection~\cite{KimSLJCKY20}. OTTO dataset contains 61,878 examples, and 93 classes, and belongs to the E-commerce domain. SNSR dataset contains 58,509 examples and 48 classes, and belongs to the Electric Currents domain. Tiny-ImageNet dataset~\footnote{https://tiny-imagenet.herokuapp.com/}, which is drawn from the Imagenet ILSVRC 2012 dataset~\cite{RussakovskyDSKS15}, has a total of 200 classes with 500 images each class for training and 50 images each class for testing. ImageNet~\cite{DengDSLL009} consists of 1000 classes, which includes 1.28 million training images. Specifically, we set $C=100$ for Tiny-ImageNet and $C=800$ for ImageNet. The streaming data simulation and other process follow the settings in the main body. 

Table \ref{tab:tab1} and Table \ref{tab:tab3}  compare the detection performance of CILF with all baseline methods on each streaming data under the single novel class and the multiple novel class cases respectively. Table \ref{tab:tab2} and Table \ref{tab:tab4} compared the forgetting performance of CILF with all baseline methods on each streaming data under the single novel class and the multiple novel class cases respectively. The results validate that CILF can still perform superior to comparing methods on detection task and with less forgetting on other domain datasets and challenging datasets.

\section{Visualization}
Figure \ref{fig:f3} shows feature embedding results within each time window using T-SNE~\cite{maaten2008visualizing}, in which each class consists of  800 randomly sampled instances. Note that we turn to utilize more complex dataset CIFAR-10 as an example, rather than simpler MNIST dataset in previous methods. Clearly, the optimal discriminative method (CPE) and generative method (DEC) are greatly interfered by the embedding confusion. Meanwhile, the output of CILF displays clearer distinctions among groups compared to other methods, which is attributed to the prototype based loss for model training. Moreover, instances from novel classes are well separated from known clusters, which is benefit for novel class detection. The compactness of new class indicates the effectiveness of curriculum clustering operator, in which reliable prototypes are developed and the model is fine-tuned from easy to difficult. The key of T-SNE is to convert the distance between high dimensional data points into Gaussian distribution probability, which can maintain the local and global similarities. However, the drawbacks of T-SNE are the high computation cost and non-convex optimization~\cite{maaten2008visualizing}. Thereby, we randomly sample 800 instances from each class following~\cite{WangKCTK19}. However, the model incremental update leads to two phenomena: 1) the embedding representation of the same example differs at different time periods; 2) the parameters of the deep network also differ in different time periods. Meanwhile, T-SNE has randomness in the optimization process~\cite{maaten2008visualizing}. Thereby, embedding varies for a certain class across the time-windows. This phenomenon also occurs in Figure 4 in~\cite{WangKCTK19}, the embedding of a certain class varies across different models. On the other hand, though the embedding varies for a certain class across the time-windows, the inter-class and intra-class relationships become clearer with the model update, which is more conducive to classification. 

Figure \ref{fig:f5} shows feature embedding results within five time window using T-SNE for multiple novel class case, in which each class consists of 800 randomly sampled instances. Similarly, the output of CILF has more distinct groups from different classes compared to other methods, which indicates that CILF can solve the embedding confusion effectively.



\ifCLASSOPTIONcompsoc

\ifCLASSOPTIONcaptionsoff
  \newpage
\fi

\bibliographystyle{IEEEtranN}{\small
\bibliography{CILF}}


